\documentclass[12pt]{iopart}

\pdfoutput=1
\usepackage{hyperref}  
\usepackage{graphicx}
\usepackage{newtxtext}
\usepackage{wasysym}
\usepackage{latexsym}
\usepackage{breqn}
\usepackage{bm}
\usepackage{cite}
\usepackage{bm}
\usepackage{fancyhdr}
\usepackage{wasysym}
\usepackage{datetime}
\usepackage{caption}
\usepackage{subcaption}
\usepackage[symbol]{footmisc}
\newcommand{\angstrom}{\mbox{\normalfont\AA}}

\newcommand{\NbSn}{Nb\textsubscript3Sn}
\newcommand{\MgB}{MgB\textsubscript2}

\newcommand{\Zs}{$Z$}
\newcommand{\YBCO}{YBa\textsubscript{2}Cu\textsubscript{3}O\textsubscript{7-$\delta$}}

\newcommand{\iu}{\mathrm{i}\mkern1mu} 
\newcommand{\mw}{$\mu w$}

\usepackage{color}
\usepackage{soul}
\usepackage[normalem]{ulem}

\graphicspath{{./Figures/}}
\begin{document}

\title[$\mu$w measurements of high field vortex parameters in \NbSn]{Microwave measurements of the high magnetic field vortex motion pinning parameters in \NbSn}

\author{Andrea~Alimenti\textsuperscript{1}, Nicola~Pompeo\textsuperscript{1}, Kostiantyn~Torokhtii\textsuperscript{1}, Tiziana~Spina\textsuperscript{2}, Ren\'{e}~Fl\"{u}kiger\textsuperscript{3}, Luigi~Muzzi\textsuperscript{4}, Enrico~Silva\textsuperscript{1}}

\address{\textsuperscript{1}Universit\`a Roma Tre, Department of Engineering, 00146 Roma, Italy}
\address{\textsuperscript{2}Fermi National Accelerator Laboratory, Batavia, IL, 60510, USA}
\address{\textsuperscript{3}University of Geneva, Department of Quantum Matter Physics (DQMP), Geneva, Switzerland}
\address{\textsuperscript{4}ENEA, 00044 Frascati (RM), Italy}
\ead{andrea.alimenti@uniroma3.it}
\vspace{10pt}

\begin{abstract}
The high frequency vortex motion in \NbSn\ was analyzed in this work up to 12~T. We used a dielectric loaded resonator tuned at 15~GHz to evaluate the surface impedance \Zs\ of a \NbSn\ bulk sample (24.8 at.\%Sn). From the field induced variation of \Zs, the high frequency vortex parameters (the pinning constant $k_p$, the depinning frequency $\nu_p$ and the flux flow resistivity $\rho_{ff}$) were obtained over a large temperature and field range; their field and temperature {dependences} were analyzed.
Comparison with other superconducting materials shows that high frequency applications in strong magnetic fields are also feasible with \NbSn. 
In the present work, we report the first measurements about the microwave response in \NbSn\ in strong magnetic fields.
\end{abstract}

%
%
%
%
%

\section{Introduction}

Among superconducting (SC) materials,  \NbSn\ is currently one of the most used in technological applications due to its interesting superconductive and mechanical properties. Despite being a well-known material, new perspective applications of \NbSn, such as superconductive radio frequency  cavities (SRFC) \cite{posen2015proof,lee2020grain} also for magnetic environments \cite{shokair2014future,di2019microwave,alesini2019galactic,braine2020superconducting} and improved magnets for new particles accelerators (e.g. the High Luminosity upgrade of LHC, or the Future Circular Collider FCC \cite{ballarino2015targets,tommasini201616}), are revamping the interest in \NbSn\ characterization \cite{pudasaini2020analysis,fernandez2019characterization,keckert2018surface}. In fact, it is necessary to test this SC in the new challenging working conditions of these applications, to better understand how to improve its performances. 

In particular, the interest in \NbSn\ is increasing, the goal being to improve the high frequency performances in view of its use in SRFC particle accelerators. 
At present, the most used material for this application field is elementary Nb. However, the need to improve the performances of RF cavities and to achieve higher accelerating fields motivated the search for a new material.
\NbSn\ is a good candidate for this application but there is still a need to study why the predicted performances (i.e. superheating field) are still far from those experimentally obtained \cite{transtrum2011superheating,liarte2017theoretical,posen2017nb3sn,posen2015radio,trenikhina2017performance}. Local geometrical surface defects are often identified as being responsible for these low performances. {A recent theoretical study has identified in the broadening of the density of states, in regions with higher pair-breaking scattering rates, a source of local heating and thus of decrease of the superheating field  \cite{Kubo2020}.}

As reported above, different kinds of RF-cavities are expected to work in presence of moderate to high static magnetic fields. 
It its well known that superconductors, at high frequencies and in the presence of magnetic fields, can exhibit surface resistances comparable to those of normal conductors.
In fact, under these conditions the main dissipative phenomenon is related to the vortex oscillations induced by the impinging electromagnetic (e.m.) wave.
For these applications, materials are searched with properties being quite different from those needed for the realization of standard SRFC cavities. Indeed, cavities optimized for zero static magnetic field require a pure superconductor with ideally no pinning centers to completely remove the trapped field after cooldown. However, in finite static magnetic fields, strong pinning
is needed to avoid large oscillations of the fluxons. In particular, above the so called depinning frequency $\nu_p$ the vortices move in the  highly dissipative flux-flow regime \cite{gittleman1966radio,silva2017vortices,PompeoLTP2020}. Hence, the measurement of $\nu_p$ in high magnetic fields is a discriminating parameter for the application of SC materials in experiments in dc magnetic fields.

In many power applications of \NbSn, the knowledge of vortex pinning is essential. Even if  microwave (\mw) measurements 
do not directly yield design parameters for dc applications, they provide useful information about the pinning characteristics, in addition to those obtained by the dc characterization techniques. A better comprehension of the pinning phenomenon is only obtained by merging 
the different information given by different dynamical regimes \cite{7433952} and \mw\ can help to unveil new vortex pinning regimes \cite{pompeo2013anisotropy}. 
Many aspects of the physics of \NbSn\ have been already studied.
For what concerns the high frequency regime, \NbSn\ surface impedance \Zs\  measurements were performed in the  {1-10~GHz range and allowed to observe deviations of the measured \Zs\ from the BCS theory and a particularly large gap $\Delta_0/k_BT=2.15$ (being $k_B$ and $T$ the Boltzman constant and the temperature, respectively) \cite{arnolds1979surface,blaschke1980electromagnetic,blaschke1981influence}.  The higher frequency behavior (at 87~GHz) was explored in \cite{perpeet1997high} confirming the large superconductive gap in \NbSn, $1.8<\Delta_0/k_BT<2.2$.} Despite the large $\Delta_0$, caused by a strong electron-phonon coupling in \NbSn, a typical BCS signature on the conductivity temperature dependence (e.g. a large coherence peak in the real part \cite{tinkham1996introduction}) was observed at 87~GHz \cite{hein1999pair,perpeet1997high}. 
Since the experimentally determined penetration depth $\lambda$ was shown to be close to the expected BCS value $\lambda_{BCS}$ \cite{hein1999high}, the latter is often used when analyzing the experimental \NbSn\ data \cite{hein1999high}.
The first \Zs\ measurements in \NbSn\ at low magnetic fields (not larger than 12~mT) and in the non-linear region were presented in\cite{andreone1997nonlinear}. 

{As it can be seen} from the present description of the high frequency behavior of  \NbSn, no studies exist on the high frequency vortex motion regime in high static magnetic field. 
%
%
We present in this work a complete microwave (${\sim15}$~GHz) characterization  of \NbSn\ up to 12~T (with preliminary results reported in \cite{Alimenti2019}) to provide new useful information about high frequency vortex motion physics in this SC.  Thus, this work fills the gap of knowledge in the high frequency behavior of  \NbSn\ in high magnetic fields.
%
%
 In particular, the surface impedance ${Z(T,H)}$ of a bulk \NbSn\ polycristalline sample is here measured with a dielectric loaded resonator (DR) \cite{alimenti2019challenging} in zero field cooling (ZFC) condition at fixed temperature $T$, and in field cooling (FC) condition at fixed applied magnetic field $\mu_0H$ values up to 12~T. 
Then, with a classical electrodynamics approach the complex resistivity $\tilde\rho(T,H)$  is obtained and analyzed with the Coffey--Clem model \cite{coffey1991unified} in order to obtain the complex vortex motion resistivity $\tilde\rho_{vm}(T,H)$ of \NbSn. 
Assuming  negligible thermal phenomena,  $\tilde\rho_{vm}$ is only a function of the real flux flow resistivity ${\rho_{ff}}$, the depinning frequency $\nu_p$ and the measurement frequency $\nu_0$ \cite{gittleman1966radio}. {Thus, ${\rho_{ff}}$ and $\nu_p$ are obtained  resorting to literature values of the London penetration depth, which is a well known quantity in \NbSn\ \cite{kneisel1977properties,arnolds1979surface,keckert2019critical}.} The measured $\nu_p$ of bulk \NbSn\ is remarkably  high when compared with $\nu_p$ in thin Nb films. 

The measured $\rho_{ff}$ is shown to exhibit a conventional Bardeen--Stephen behavior \cite{PhysRev.140.A1197}. The scaling of the $\rho_{ff}$ with the applied magnetic field allowed us to evaluate the upper critical field $H_{c2}(T)$ down to 4~K. The so obtained $H_{c2}(T)$  is well fitted by the Maki-de~Gennes approximation \cite{maki1964magnetic,de1964behavior}, as expected from other works \cite{godeke2005upper}. 

Following \cite{pompeo2008reliable} we extended the analysis of the high frequency vortex pinning characteristics in \NbSn\ considering the contribution of the thermal creep: based on analytical constraints of the used equations and physical limits, a statistical approach is used to assess probability intervals of the evaluated pinning parameters. 


The paper is organized as follows: in Sec.~\ref{sec:theory} the high frequency vortex motion is briefly described, in Sec.~\ref{sec:method} the measurement method is presented, then the sample characteristics are reported in Sec.~\ref{sec:sample}. Finally the results are presented in Sec.~\ref{sec:results} and in Sec.~\ref{sec:Comparison} a comparison of the \mw\ performances of \NbSn\ with those of \MgB\ and \YBCO\ is performed.

\section{Surface impedance of superconductors in the mixed state}\label{sec:theory}
The surface impedance $Z$ is the complex physical quantity commonly used to describe the electromagnetic (e.m.) response of good conductors \cite{chen2004microwave}. It is defined as the ratio ${Z=E_\parallel/H_\parallel}$ \cite{jackson2007classical}, where ${E_\parallel}$ and ${H_\parallel}$  are respectively the electric and magnetic fields components parallel to the surface of the conductor. $Z$ contains interesting information about the dissipative and energy storing effects of the material under study. For bulk materials, in the local limit,  ${Z=\sqrt{\iu\omega\mu_0\tilde\rho}}$ \cite{jackson2007classical}, where $\omega=2\pi\nu$ is the angular frequency of the impinging  e.m. wave, $\mu_0$ is the vacuum magnetic permeability and $\tilde\rho$ is the complex resistivity of the material. Since in this work we deal with a \NbSn\ superconductive bulk sample in high magnetic field, $\tilde\rho$ contains both {the super/normal fluid complex charge transport contributions} and the vortex flow characteristics as presented in \cite{coffey1991unified}. The first contribution is modeled by the two-fluid conductivity ${\sigma_{2f}=\sigma_1-\iu\sigma_2}$  and the second by the complex vortex motion resistivity $\rho_{vm}$, thus ${\tilde\rho=f\left(\sigma_{2f},\rho_{vm}\right)}$. Far enough from the superconductive transition, where ${\sigma_2\gg\sigma_1}$, the normal fluid screening effect is weak enough to be neglected and $Z$ is written as:
\begin{equation}\label{eqn:Zrho}
Z\simeq\sqrt{\omega\mu_0\left(-\frac{1}{\sigma_2}+\iu\rho_{vm}\right)};
\end{equation} 
where ${\sigma_2=1/\omega\mu_0\lambda^2}$, with $\lambda$ the London penetration depth. When no external magnetic field is applied ${\rho_{vm}=0}$ and ${Z\simeq\sqrt{-\omega\mu_0/\sigma_2}=\iu\omega\mu_0\lambda}$.

With high frequencies (microwaves) excitation and low e.m. field amplitude the vortices start oscillating around their equilibrium positions (the pinning centres) as damped harmonic oscillators and their dissipative and reactive response depends on the pinning potential characteristics. Within the harmonic oscillator formalism, we can describe the Lorentz force due to the interaction between the microwave induced currents $\bm{J}_{\mu w}$ and the magnetic flux quanta $\bm{\Phi}_0$ as the driving force, the pinning effect  as a linear elastic force $\bm{F}_{p}=-k_p\bm{x}$ (small oscillation), the non-equilibrium conversion between quasi-particles and condensate during the fluxons motion as a dissipative viscous drag force $\bm{F}_{drag}=-\eta\bm{v}$ and finally the thermal creep as a stochastic thermal force $\bm{F}_{th}$ \cite{coffey1991unified,gittleman1966radio,pompeo2008reliable,tinkham1996introduction, silva2017vortices, PompeoLTP2020}. Thus, assuming massless fluxons \cite{Kopnin2002} the dynamic equation of motion becomes:
\begin{equation}
\bm{J}_{\mu w}\times\bm{\Phi}_0+\bm{F}_{th}=k_p\bm{x}+\eta\bm{v}\;,
\end{equation}
with $k_p$ the pinning constant, $\bm{x}$ the fluxon displacement, $\eta$ the viscous drag coefficient and $\bm{v}$ the fluxon velocity.
The Coffey-Clem (CC) vortex motion resistivity ${\rho_{vm}}$ is then obtained \cite{coffey1991unified,pompeo2008reliable}:
\begin{equation}\label{eqn:rhoVM}
\rho_{vm}=\rho_{ff}\frac{\epsilon+\iu\nu/\nu_c}{1+\iu\nu/\nu_c}\;,
\end{equation}
where ${\rho_{ff}=\Phi_0B/\eta}$ is the flux-flow resistivity, $B$ the magnetic flux density. In the London limit $B\simeq\mu_0H$, with $H$ the applied magnetic field strength. The thermal creep contribution is taken into account by the adimensional creep factor $0\leq\epsilon\leq1$  \cite{coffey1991unified,pompeo2008reliable}. $\nu_c$ is the characteristic frequency of the vortex motion, marking the crossover between an elastic vortex motion regime ($\nu\ll\nu_c$) and an highly dissipative regime ($\nu\gg\nu_c$). When $\epsilon\rightarrow0$, no  flux creep exists and $\nu_c\rightarrow\nu_p$ with $\nu_p$ the depinning frequency, defined as $\nu_p=k_p/(2\pi\eta)$. In the case of small oscillations here relevant, $k_p$ is the pinning linear elastic constant which for rigid fluxons is a measure of the pinning well steepness \cite{pompeo2008reliable,golosovsky1996high}. The $\epsilon\rightarrow0$ limit is known in literature as the Gittleman-Rosenblum (GR) model \cite{gittleman1966radio}:
\begin{equation}\label{eqn:rhoGR}
\rho_{vm,GR}=\frac{\Phi_0 B}{\eta}\frac{1}{1-\iu\frac{\nu_p}{\nu}}\;.
\end{equation}
 In the high creep limit $\epsilon\rightarrow1$ the fluxons behave as free fluxons due to thermal jumps and a free-flux flow regime takes place. The $\epsilon(U_0)$ and $\nu_c(\nu_p,U_0)$ dependences on the creep activation energy $U_0$ depend on the pinning potential shape and thus on the particular model used to describe the pinning profile \cite{coffey1991unified,brandt1992linear}.   

Microwave measurements are particularly versatile since they allow to obtain a measure of both the pinning shape/steepness $k_p$ and of the free flux-flow resistivity $\rho_{ff}$ which, particularly for bulk samples as in this case, would  require high dc-current to be properly measured.
In the following we describe how we obtain the vortex motion parameters in \NbSn\ bulks.

\section{Measurement system and method}\label{sec:method}
In this section we briefly outline how $Z$ is obtained with our measuring system based on a dielectric loaded resonator (DR). Further information about the measurement technique with an in-depth uncertainties analysis is reported in \cite{alimenti2019challenging}.

\begin{figure}[htbp]
\centering
\includegraphics[width=0.5\textwidth]{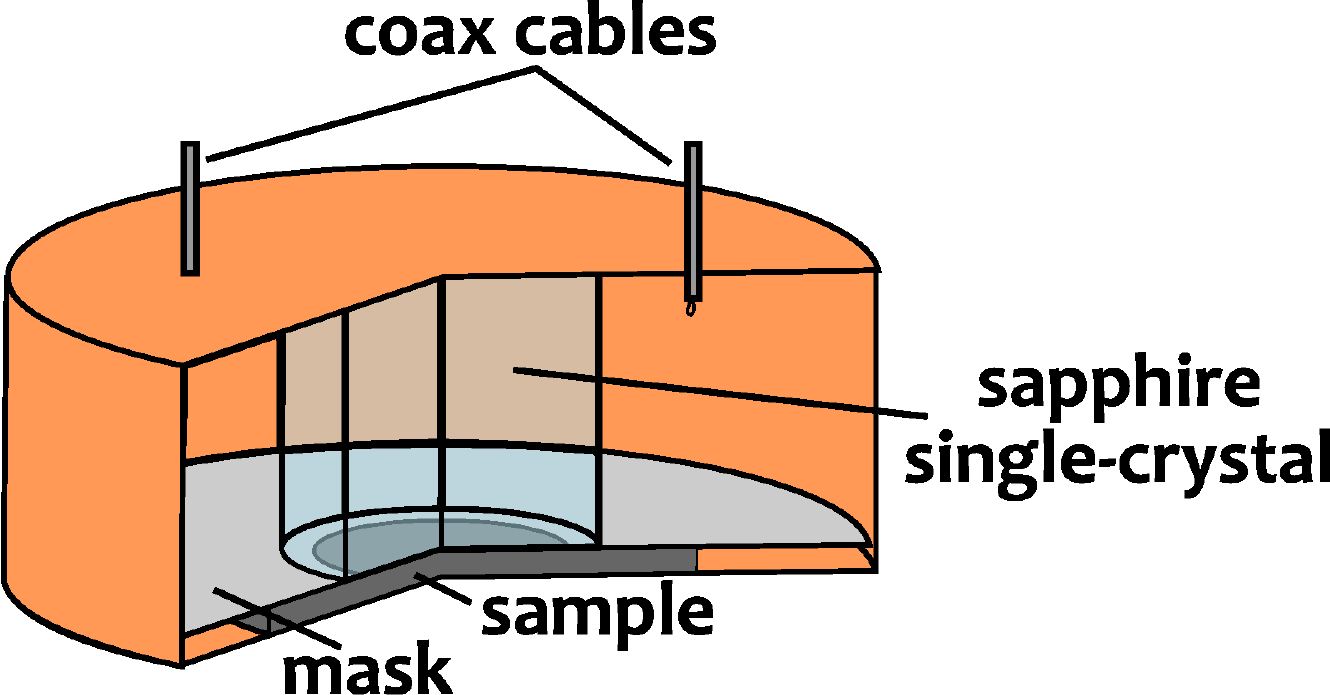}
\caption{Sketch of the dielectric loaded resonator.}
\label{fig:DR}
\end{figure}

DRs are a measurement standard for superconductors $Z$ characterization (IEC 61788-7:2020)  \cite{IECstandard} thanks to their high sensitivity. In the IEC standard twin SC samples are used and both resonator bases are covered by a SC sample. As shown in \cite{alimenti2019challenging} the double sample configuration is not always the right choice for the high field measurements, because of a lack of sensitivity in presence of high dissipations. For this reason, in this case, the single sample configuration (see \figurename~\ref{fig:DR}) provides better performances \cite{alimenti2019challenging}. Our set-up is composed by a copper cylindrical cavity loaded with a sapphire crystal as represented in \figurename~\ref{fig:DR}. The choice of the metallic enclosure is forced by the need to perform measurements in high magnetic fields: the use of a SC cavity would add a magnetic background of difficult evaluation.
The used dielectric is a  single-crystal sapphire cylinder ($h=5.00$~mm, $\diameter=8.00$~mm). The relatively high dielectric permittivity ($\varepsilon_\parallel\sim11.5$, $\varepsilon_\perp\sim9.5$) and the low losses of sapphire are used to increase the measurement sensitivity by reducing the conduction losses on the lateral wall of the resonator \cite{alimenti2019challenging}. 

An Anritsu Vector Network Analyzer (VNA) 37269D, is used to measure the four complex scattering parameters of the DR. The VNA and the DR are linked through a cryogenic and non-magnetic K-type coaxial transmission line. The resonator is excited in the TE\textsubscript{011} resonating mode at $\sim15$~GHz, and it is characterized in transmission. The acquired scattering parameters are then fitted with a modified Fano resonance curve \cite{petersan1998measurement,pompeo2017fitting} which is used to obtain the unloaded quality factor $Q$ and the resonance frequency $\nu_{0}$ of the resonator. Measurements are performed at low \mw\ power levels, the peak RF magnetic field impinging on the surface of the sample is estimated to be ${<10\;\mu\mbox{T}}$, to characterize  $Z$ in the linear regime. We did not observe any power dependence of the response in the range of temperature, fields and power level here explored.

The sample under study is loaded into the cavity in order to substitute a base of the resonator (end-wall perturbation method) and covered with a planar metallic mask, with a central circular hole ($\diameter\sim6$~mm), to preserve the cylindrical geometry.
%
%

When an external magnetic field $H$ is applied at a temperature $T'$ the variation $\Delta Z(T',H)=Z(T',H)-Z(T',0)$ of $Z$   is obtained as follows :
\begin{equation}\label{eqn:DZ}
\Delta Z(T',H)=G_s\Delta\frac{1}{Q(T',H)}-2\iu G_s\frac{\Delta\nu_{0}(T',H)}{\nu_{ref}}- \Delta bckg(T',H);
\end{equation}
 where $G_s\approx2700\;\Omega$ is a geometrical factor evaluated with electromagnetic simulations and $\Delta x(T',H)$ indicates a variation of $x(T',H)$ parameter with respect to the reference value obtained with no applied magnetic field $x(T',0)$. $\nu_{ref}$ is the reference resonance frequency at $H=0$~T. Finally, $bckg$ is a complex parameter which represents the response of the resonator itself. Since both the DR and the measurement system were carefully designed to operate in high magnetic fields, $bckg$ is very weakly field dependent \cite{alimenti2019challenging}, with respect to the SC sample variation $\Delta Z(T',H)$, thus we assume $\Delta bckg(T',H)\sim0$ \footnote{The weak magnetic contribution of the copper resonator was evaluated up to 12~T in the homogeneous configuration (without any SC sample loaded) to be: {$\Delta \nu_0<2.5$~kHz~T\textsuperscript{-1} and $\Delta Q<40$~T\textsuperscript{-1} with $\nu_0\sim15$~GHz and $Q\sim17900$.}}. 

{Field cooling (FC) and zero field cooling (ZFC) measurements were performed and they are discussed in the next Section. In FC condition the magnetic field was applied before cooling. 
After cooling down to $\sim6$~K, the temperature was raised at a constant rate 0.1~K/min.
In ZFC the sample was cooled without an externally applied magnetic field; when the target temperature was reached and stabilized within $\pm0.05$~K, the magnetic field was swept at 0.3~T/min up to 12~T then down to -12~T and back to 0~T and the reversible component isolated.}

\section{The sample}\label{sec:sample}
The \NbSn\ sample platelet was obtained starting from a polycrystalline bulk piece sintered by Hot Isostatic Pressure (HIP) technique (2~kbar Argon pressure at $1250\;^{\circ}$C for 24h) at the University of Geneva \cite{spina2015proton}. After HIP, the \NbSn\ bulk piece was cut into tiny platelets by means of spark erosion and each platelet was then polished with SiC grinding papers and submitted to "flash-anneal" heat treatment {($900\;^{\circ}$C/10~min)} for stress release. 

Microstructural and magnetization analyses reveal an average grain size of $\sim20\;\mu{\mbox{m}}$, a composition very close to stoichiometry (24.8 at.\%Sn) and a sharp superconducting transition at 17.9~K reflecting the high quality and homogeneity of these samples. Finally, from Rietveld refinement the lattice constant and the Bragg-Williams long-range order parameter have been estimated to be 5.291~$\angstrom$ and 0.98 respectively. Further details on the procedure and analysis can be found in \cite{flukiger2017variation}.
A sample of approximate area of 30 mm$^2$ was chosen for the present study.

\section{Results}\label{sec:results}
In this section we first show the $Q$ and $\nu_0$ measurements to check the calibration process through the comparison of the obtained normal state sample characteristics with literature values. Then, we derive the vortex parameters under the common assumption of negligible thermal creep (i.e. Gittleman-Rosenblum (GR) model \cite{gittleman1966radio}): the use of the GR model is the standard analysis procedure \cite{golosovsky1996high, Maeda2005, silva2017vortices, PompeoLTP2020} so that it allows to easily compare the results on \NbSn\ with other materials. Finally, in the last subsection, the contribution given by the flux creep is evaluated with a statistical analysis of the obtained data.


\subsection{Normal state}\label{sec:NormalS}
\begin{figure}[htbp]
\centering
\begin{subfigure}[t]{1\linewidth}
\centering
\includegraphics[width=0.6\linewidth]{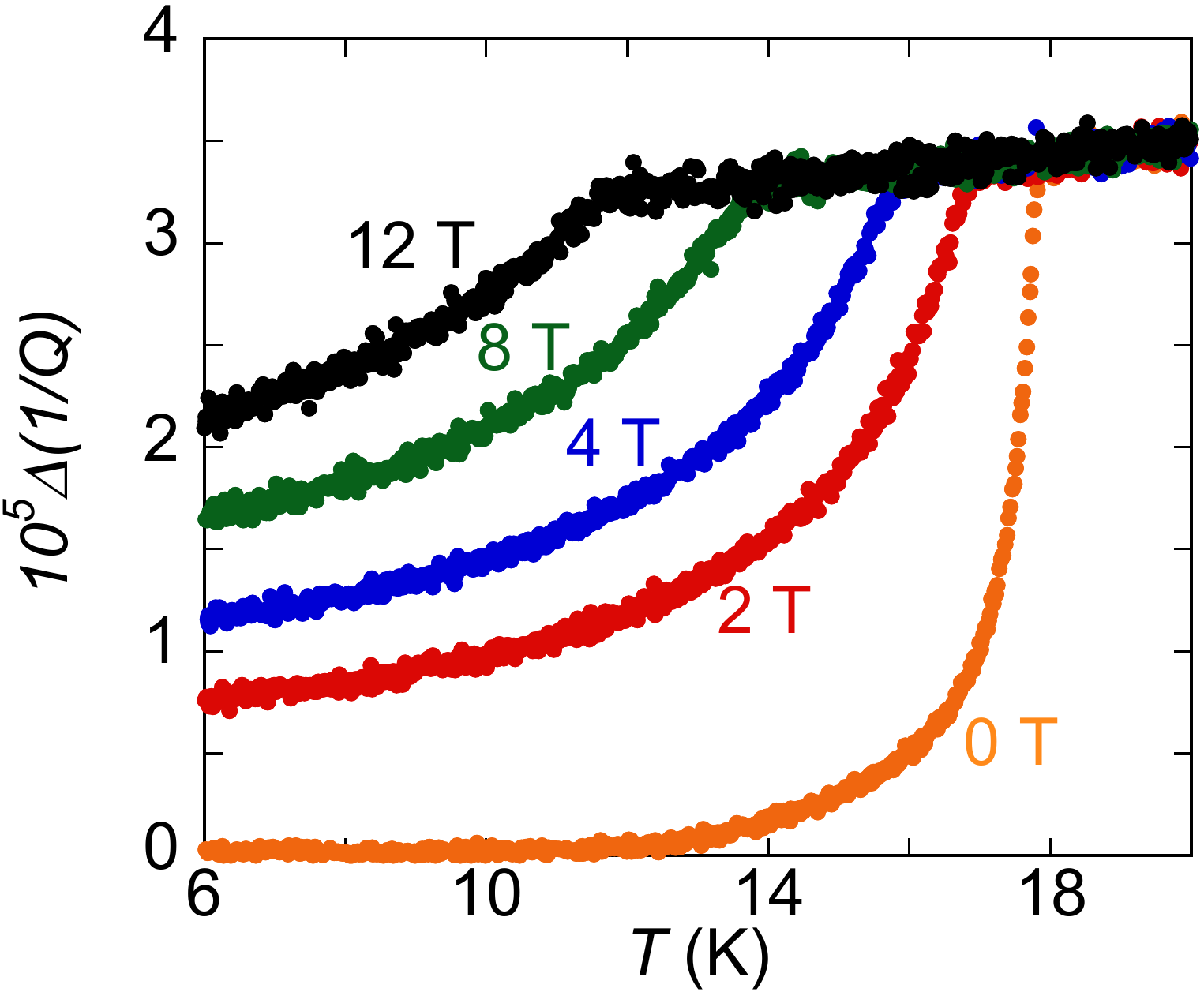}
\caption{}
\label{fig:D1Q}
\end{subfigure}
\begin{subfigure}[t]{1\textwidth}
\centering
\includegraphics[width=0.6\linewidth]{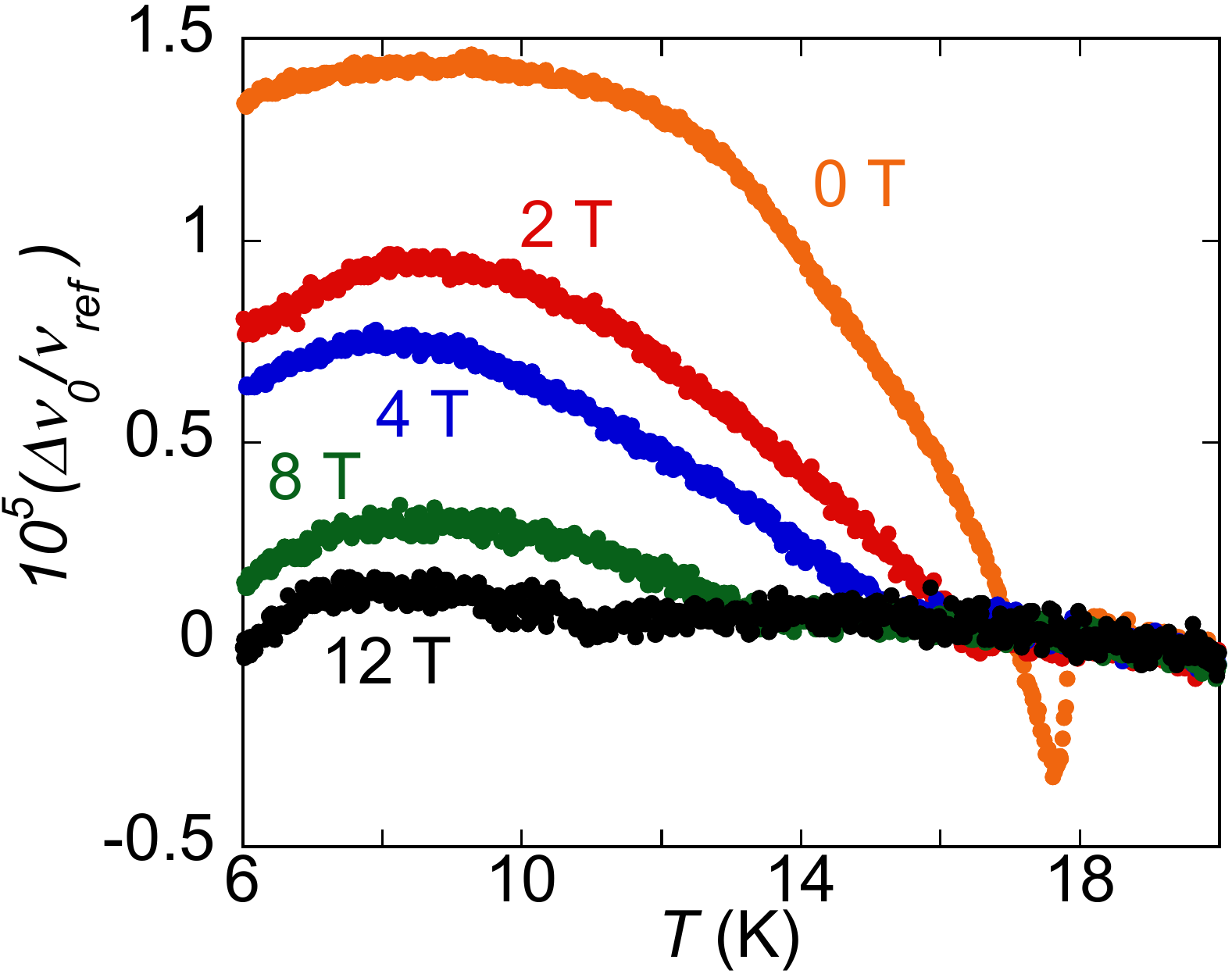}
\caption{}
\label{fig:Df0}
\end{subfigure}
\caption{(a) Variation ${\Delta(1/Q)=Q(T,H)^{-1}-Q(T\rightarrow0,H=0)^{-1}}$ measured in field cooling condition at {different fields (i.e. 0~T, 2~T, 4~T, 8~T, 12~T)}. (b) Variation ${\Delta \nu_0/\nu_{ref}=(\nu_0(T,H)-\nu_{ref})/\nu_{ref}}$ with ${\nu_{ref}=\nu_0(T=T_c)}$ measured in field cooling condition at {different fields (i.e. 0~T, 2~T, 4~T, 8~T, 12~T)}. The temperature background of the resonator, which gives rise to the $\nu_0$ hump, is evident requiring the calibration procedure described in the text.}
\label{fig:DQDf}
\end{figure}

In \figurename~\ref{fig:DQDf}  we show the measured variation ${\Delta(1/Q)=Q(T,H)^{-1}-Q(T\rightarrow0,H=0)^{-1}}$ and ${\Delta \nu_0/\nu_{ref}=(\nu_0(T,H)-\nu_{ref})/\nu_{ref}}$ with ${\nu_{ref}=\nu_0(T=T_c)}$ at {$\mu_0H_a=\{0,\,2,\,4,\,8,\,12\}$~T} in FC condition. Since below $\sim20$~K the copper and sapphire losses do not depend on the temperature  the height of the $Q^{-1}$ transition (\figurename~\ref{fig:D1Q}) can be assigned to $\Delta R$ of \NbSn. Thus, from Eq.~(\ref{eqn:DZ})  ${R_n=94.6}$~m$\Omega$ with $R_n$ the normal state surface resistance. From $R_n$, the normal state resistivity is obtained from the normal skin effect as ${\rho_n=2R_n^2/\omega\mu_0=14.8\;\mu\Omega\mbox{cm}}$.
%



An estimation of $\rho_n$ based on the long range order parameter $S$, yields ${\rho_n=147(1-S^4)\;\mu\Omega\mbox{cm}}$ \cite{flukiger1987effect}. The measured $\rho_n$ corresponds to $S=0.97$, perfectly in agreement with the measurement obtained with X-ray diffractometry on a sample from the same batch of our platelet \cite{spina2015proton}.  

Moreover, the obtained $\rho_n$ matches well also with the atomic Sn content $\beta$ of the sample, since with ${\rho_n=14.8\;\mu\Omega\mbox{cm}}$ and from \cite{godeke2006review}, $\beta=0.25$ to be compared to our data $\beta=0.248$.  
This excellent agreement between the measured $\rho_n$ and the microscopic parameters measured on the sample from the same batch as ours represents a validation of the $G_s$ estimation. {Moreover, the composition of the bulk sample is also in agreement with the measured $T_c$ and the generally accepted ${T_c(\beta)}$ relation presented in \cite{devantay1981physical}. This confirms that the Devantay data set is more descriptive of \NbSn\ bulk samples behavior than Moore's \cite{moore1979energy} as discussed in \cite{godeke2006review}.}


\subsection{Microwave vortex motion in \NbSn}
\begin{figure}[htbp]
\centering
\begin{subfigure}[t]{1\linewidth}
\centering
\includegraphics[width=0.6\linewidth]{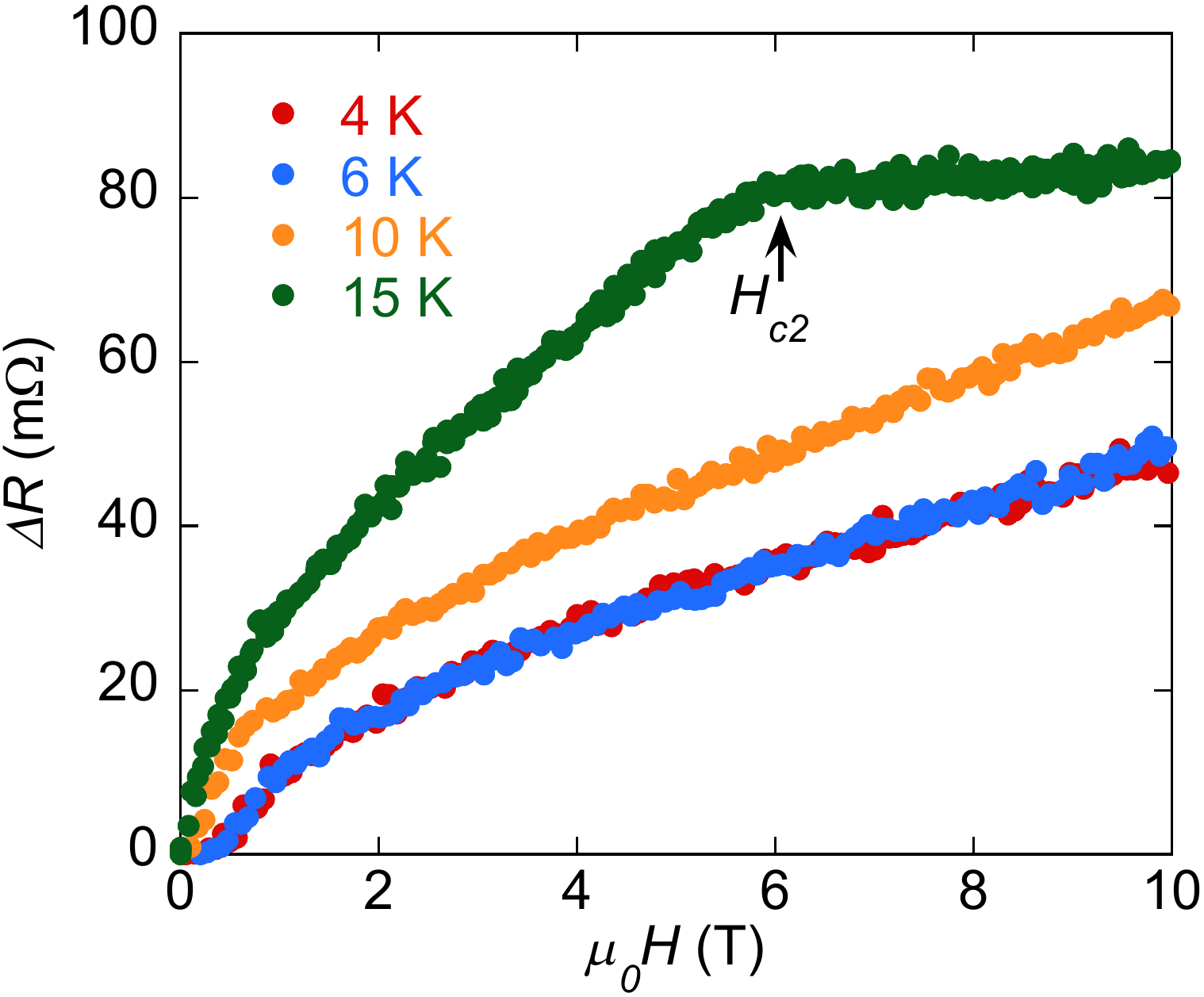}
\caption{}
\end{subfigure}
\begin{subfigure}[t]{1\textwidth}
\centering
\includegraphics[width=0.6\linewidth]{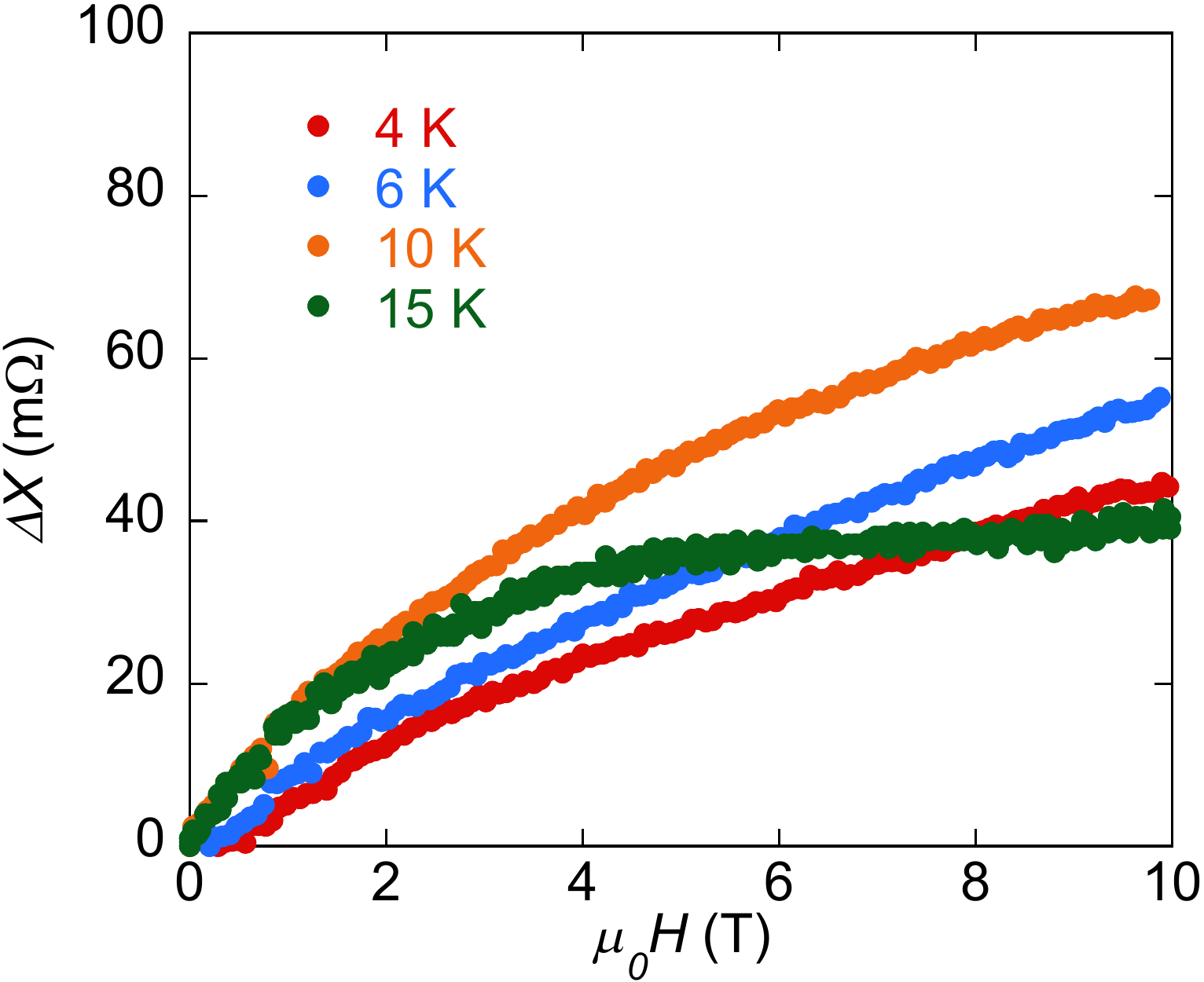}
\caption{}
\end{subfigure}
\caption{Surface impedance variation ${\Delta Z=Z(H)-Z(H=0)}$ measured in ZFC conditions at {different temperatures (i.e. 4~K, 6~K, 10~K, 15~K)}. In (a) the surface resistance $\Delta R$ and in (b) the surface reactance $\Delta X$. The knee in the 15~K $\Delta R$ curve corresponds to the superconductive phase transition: above that point the applied magnetic field $\mu_0H>H_{c2}$ and $\Delta R$ is almost flat as expected in the normal state. On $\Delta X$ the transition is smoother thus less evident.}
\label{fig:DQDf_rampe}
\end{figure}

In order to isolate the fluxon motion response of \NbSn, Eq.~(\ref{eqn:DZ}) is applied to the data shown in \figurename~\ref{fig:DQDf}. In this way the temperature background contribution, which is particularly evident in $\nu_0$ measurements (\figurename s~\ref{fig:Df0}), is removed. The same procedure is followed for the ZFC measurements shown in \figurename~\ref{fig:DQDf_rampe}. {In this case the variations of  $Q$ and $\nu_0$} are directly related to the sample $\Delta Z$ since the resonator is made only with non magnetic materials.
%
Then,  the vortex motion resistivity ${\rho_{vm}=\rho_{vm}'+\iu\rho_{vm}''}$ is obtained from the measured $\Delta Z$, using Eq.~(\ref{eqn:Zrho}), as:
\begin{equation}\label{eqn:rhovm1}
\rho_{vm}'(T,H)=2\Delta R(T,H)\left(\lambda(T,0)+\frac{\Delta X(T,H)}{\mu_0\omega}\right)\;,
\end{equation}
\begin{equation}\label{eqn:rhovm2}
\rho_{vm}''(T,H)=\frac{-\Delta R(T,H)^2+\left(\Delta X(T,H)+\lambda(T,0)\mu_0\omega\right)^2}{\mu_0\omega}-\mu\omega\lambda(T,H)^2\;.
\end{equation}
In order to reliably obtain $\rho_{vm}$, we calculate $\lambda$ from the well known values, as follows. It is known \cite{hein1999high} that $\lambda$ in \NbSn\ closely follows a BCS behavior \cite{tinkham1996introduction} although stoichiometric \NbSn\ exhibits similarities to strong coupling superconductors. It is then safe, following the common habit, to describe $\lambda$ with the BCS expression in Eq.s~(\ref{eqn:rhovm1}),~(\ref{eqn:rhovm2}), with Debye temperature $\Theta_D=230$~K  \cite{junod1983heat} and  superconducting energy gap $2\Delta=3.77k_BT_c$ \cite{bosomworth1967energy}.
Finally, it must be noticed that at high fluxons densities (in practice, just above the first penetration field), and in our measurement frequency band, the main reactive contribution is given by the vortex motion, thus Eq.s~(\ref{eqn:rhovm1}),~(\ref{eqn:rhovm2}) are very weakly sensitive to $\lambda$. This was tested using as an alternative a simple two-fluid $\lambda(T)$ temperature dependence $1-t^2$ and the discrepancies  in the following analysis were well below $5\,\%$. For the field dependence we used a $1-b^4$ superfluid fraction dependence with $b=B/B_{c2}$ the reduced field. Similarly to the $T$ dependence, also the exact field dependence does not give significant changes on the final results. 

From $\rho_{vm}$, with Eq.~(\ref{eqn:rhoVM}), $\rho_{ff}(T,H)$ and $\nu_p(T,H)$ are directly obtained within the Gittleman--Rosenblum (GR) model. As previously discussed, the GR model assumes negligible thermal effects, thus it is more descriptive of the data far from the critical surface. Nevertheless, when thermal creep is not negligible, the GR model provides a lower boundary for $\nu_p$ and $k_p$ \cite{pompeo2008reliable}. Thus, despite the model simplicity, the GR model is particularly useful for an estimation of the pinning parameters, as presented in the next subsections. 

\subsubsection{Flux-flow resistivity and viscous drag coefficient.}
\begin{figure}[hbp]
\centering
\includegraphics[width=0.6\textwidth]{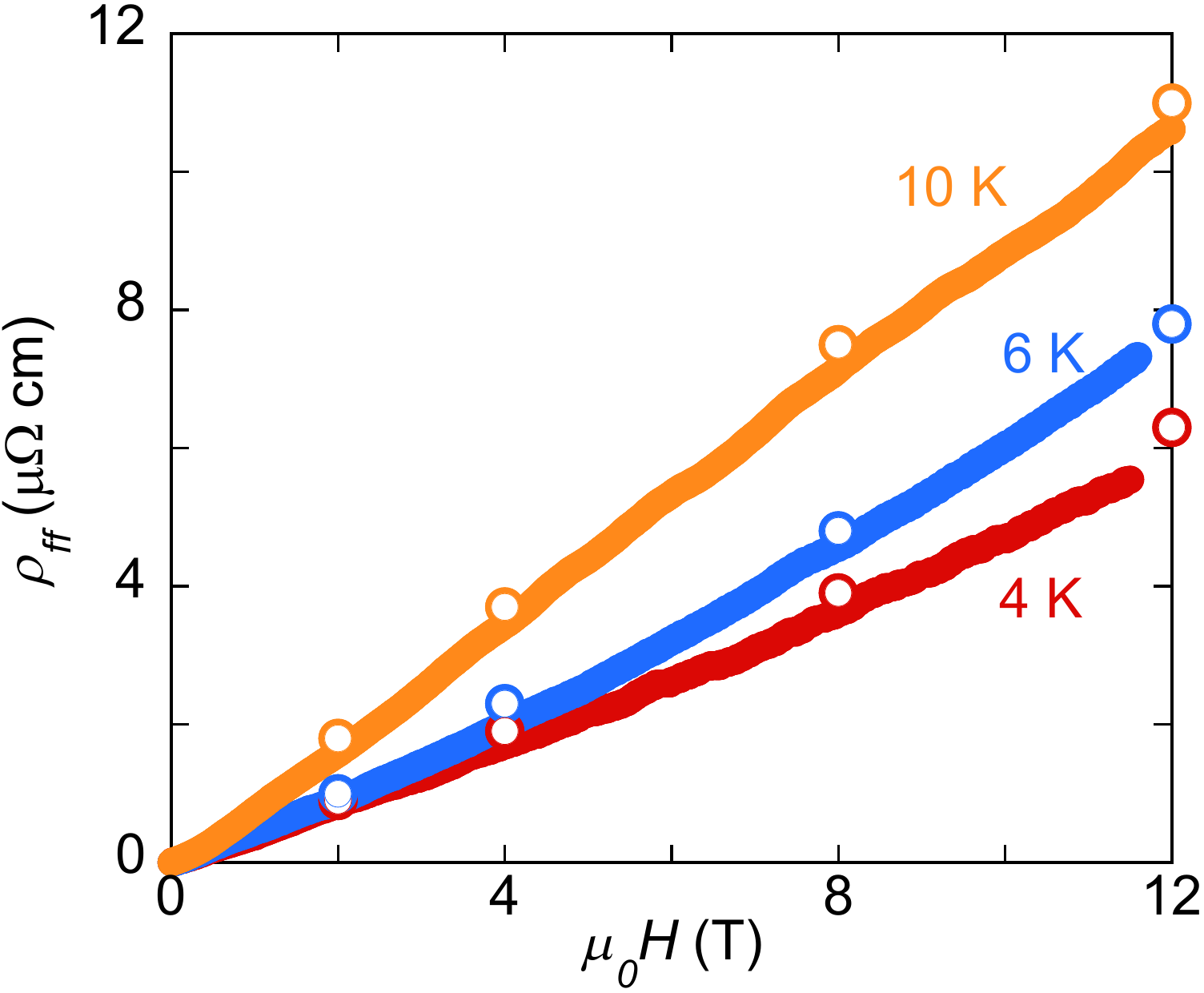}
\caption{The flux flow resistivity $\rho_{ff}$ of bulk \NbSn\ as a function of the applied magnetic field $\mu_0H\simeq B$ measured in ZFC at {different temperatures (i.e. 4~K, 6~K, 10~K)}. The sparse empty circles come from the FC measurements. }
\label{fig:rhoFF}
\end{figure}

\begin{figure}[htb]
\centering
\includegraphics[width=0.6\textwidth]{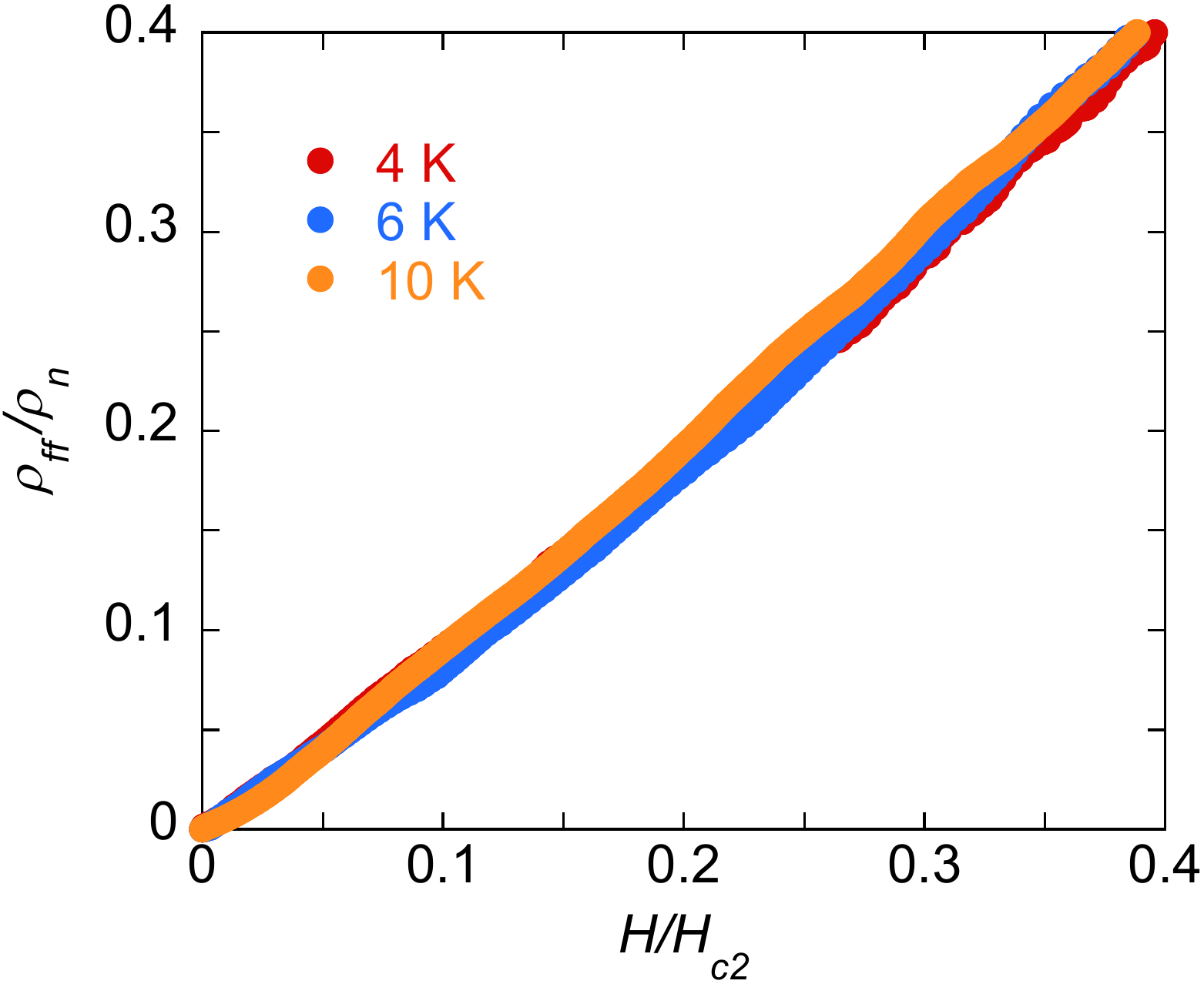}
\caption{Flux-flow resistivity $\rho_{ff}$ measured from ZFC $H$ sweeps at {4~K, 6~K, 10~K} and normalized to the normal state resistivity $\rho_n$. The field values are normalized choosing $H_{c2}(T)$ values in order to obtain unitary slope.}
\label{fig:RhoFFscaling}
\end{figure}

\begin{figure}[hb]
\centering
\includegraphics[width=0.6\textwidth]{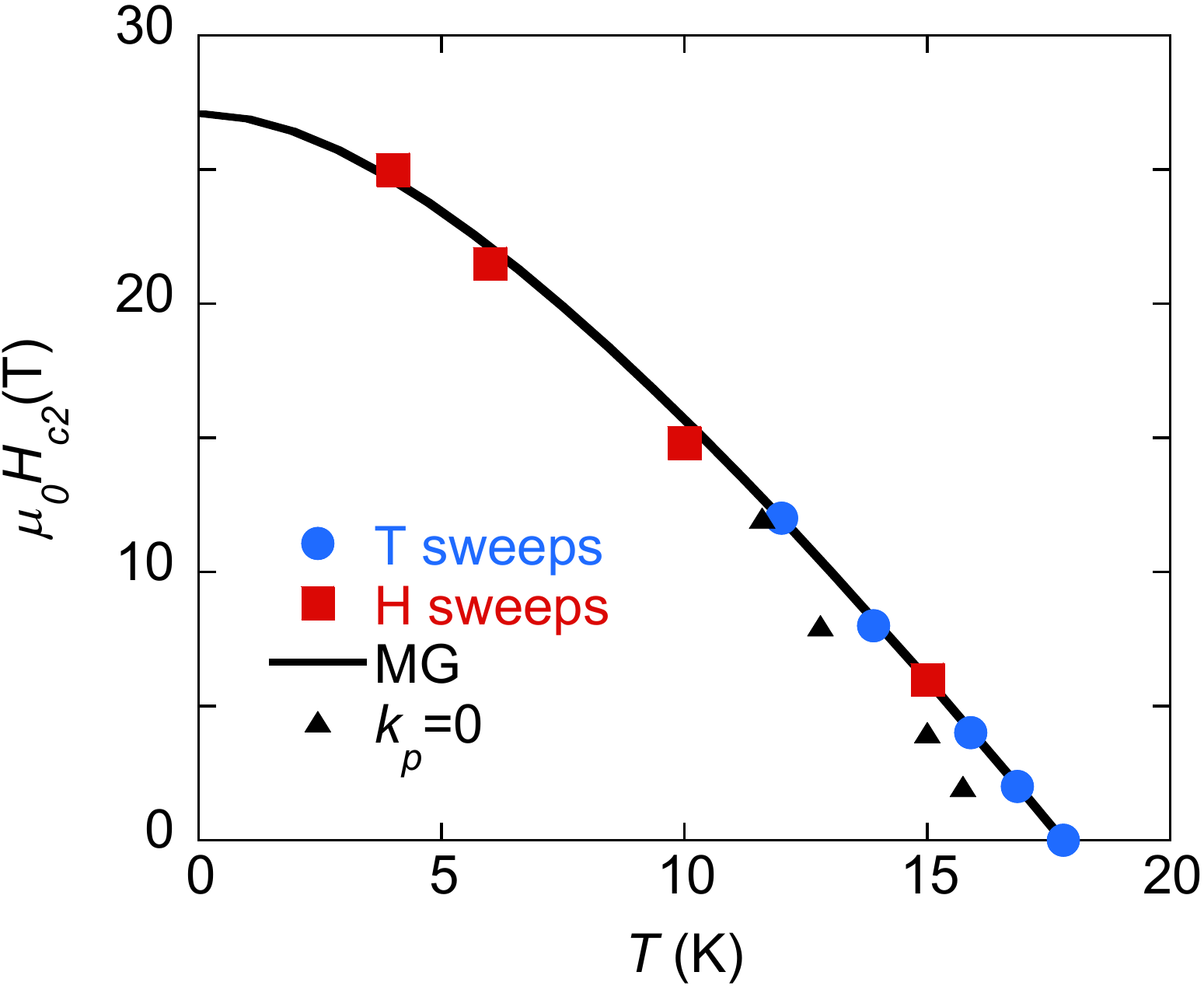}
\caption{Measured upper critical field $\mu_0H_{c2}$ temperature dependence. The full blue circles are obtained from the field cooling temperature sweeps while the full red squares from the field sweeps. The points at {4~K, 6~K, 10~K} are obtained from the $\rho_{ff}$ scaling. The data are fitted with the Maki-de Gennes model (MG) \cite{maki1964magnetic,de1964behavior}. The black triangles are obtained from the linear extrapolation of $k_p(T)$ to 0 (see Sec.~\ref{sec:pinnConst}).}
\label{fig:Bc2MG}
\end{figure}

When a vortex moves, energy is lost by the non-equilibrium conversion of the condensate in quasi-particle on the onward vortex side and the restoring of the condensate in the back side \cite{tinkham1996introduction,golosovsky1996high}. 
\figurename~\ref{fig:rhoFF} shows $\rho_{ff}$ as obtained by combining Eq.s~(\ref{eqn:rhovm1}) and (\ref{eqn:rhovm2}) and Eq.~(\ref{eqn:rhoGR}) as ${\rho_{ff}=\left({\rho_{vm}'}^2+{\rho_{vm}''}^2\right)/\rho_{vm}'}$ in ZFC. {A good overlap is found with $\rho_{ff}$ measured in FC conditions (see \figurename~\ref{fig:rhoFF}). }
The 15~K ZFC curve is not analyzed here since near the transition the unavoidable presence of flux creep prevents from performing the analysis here presented. For the same reason the derivation of the vortex parameters is estimated as possible only for $T<0.8\,T_{c2}$ where $T_{c2}=T_c(H)$.  \figurename~\ref{fig:rhoFF} correctly shows that $\rho_{ff}$ increases when $T$ and $H$ increase. This is an expected behaviour since $\rho_{ff}\propto<\tau>^{-1}$ with $\tau$ the quasiparticle scattering time in vortices core (which decreases approaching the superconductive transition \cite{tinkham1996introduction}) averaged on the Fermi surface. Moreover, \figurename~\ref{fig:rhoFF} shows an almost perfect linear behavior $\rho_{ff}\propto H$. This implies that $\eta$ is field independent in agreement with both Tinkham \cite{tinkham1996introduction} and  Bardeen-Stephen (BS) \cite{PhysRev.140.A1197} descriptions of the vortex dissipation phenomena. For $T\ll T_c$, both theories give an equivalent description of the total viscosity \cite{tinkham1996introduction, PhysRev.140.A1197}:
\begin{equation}\label{eqn:rhof_rhon}
\eta=\frac{\Phi_0B}{\rho_{ff}}\approx\frac{\Phi_0\mu_0H_{c2}}{\rho_n}\;.
\end{equation}
Equation (\ref{eqn:rhof_rhon}) allows us to scale these curves with respect to $H_{c2}$ once ${\rho_n(T)=2 R_n(T)^2/\omega\mu_0}$ is determined (Sec.~\ref{sec:NormalS}).
\figurename~\ref{fig:RhoFFscaling} shows the obtained good $\rho_{ff}$ scaling that allows a reliable determination of the upper critical field $H_{c2}(T)$ even above the maximum field reached. 
{The obtained temperature derivative ${\mu_0dH_{c2}(T)/dT|_{T_c}\simeq2.2}$~T/K and the $H_{c2}(T)$ data points which were directly observed, and/or obtained by the scaling procedure, are well fitted with the Maki-de~Gennes (MG) approximation  \cite{maki1964magnetic,de1964behavior} (see \figurename~\ref{fig:Bc2MG}) in agreement with the literature \cite{godeke2005upper}. }
The fact that the $H_{c2}$ points obtained by the scaling procedure are well placed on the MG curve further validates the use of the BS model for \NbSn\ and the scaling procedure. The MG model uses only two free parameters: $T_c$ (measured) and the normal electrons diffusion coefficient $D$. In particular, it can be shown \cite{godeke2005upper} that within the MG approximation ${\mu_0dH_{c2}(T)/dT|_{T_c}=-4\Phi_0k_B/\pi^2\hbar D}$, with $k_B$ the Boltzmann constant and $\hbar$ the reduced Planck constant. {Thus, the fit contains only experimentally determined parameters.} With the fit of the measured $H_{c2}(T)$ we obtain $D\sim5.0\times10^{-5}$~m\textsuperscript{2}/s. We point out that $H_{c2}(T)$ does not depend on the electron-phonon coupling, thus even if the simple MG approximation does not take the coupling strength into account (differently from the more complex Eliashberg theory) it can be reliably used in this case: it is shown in literature \cite{godeke2005upper} that the MG model approximates well the \NbSn\ $H_{c2}(T)$ behavior in different samples (e.g. single crystal, thin films, bulk, wires) and with different Sn contents \cite{godeke2005upper}.

\subsubsection{Pinning constant.}\label{sec:pinnConst}
\begin{figure}[htpb]
\centering
\includegraphics[width=0.7\textwidth]{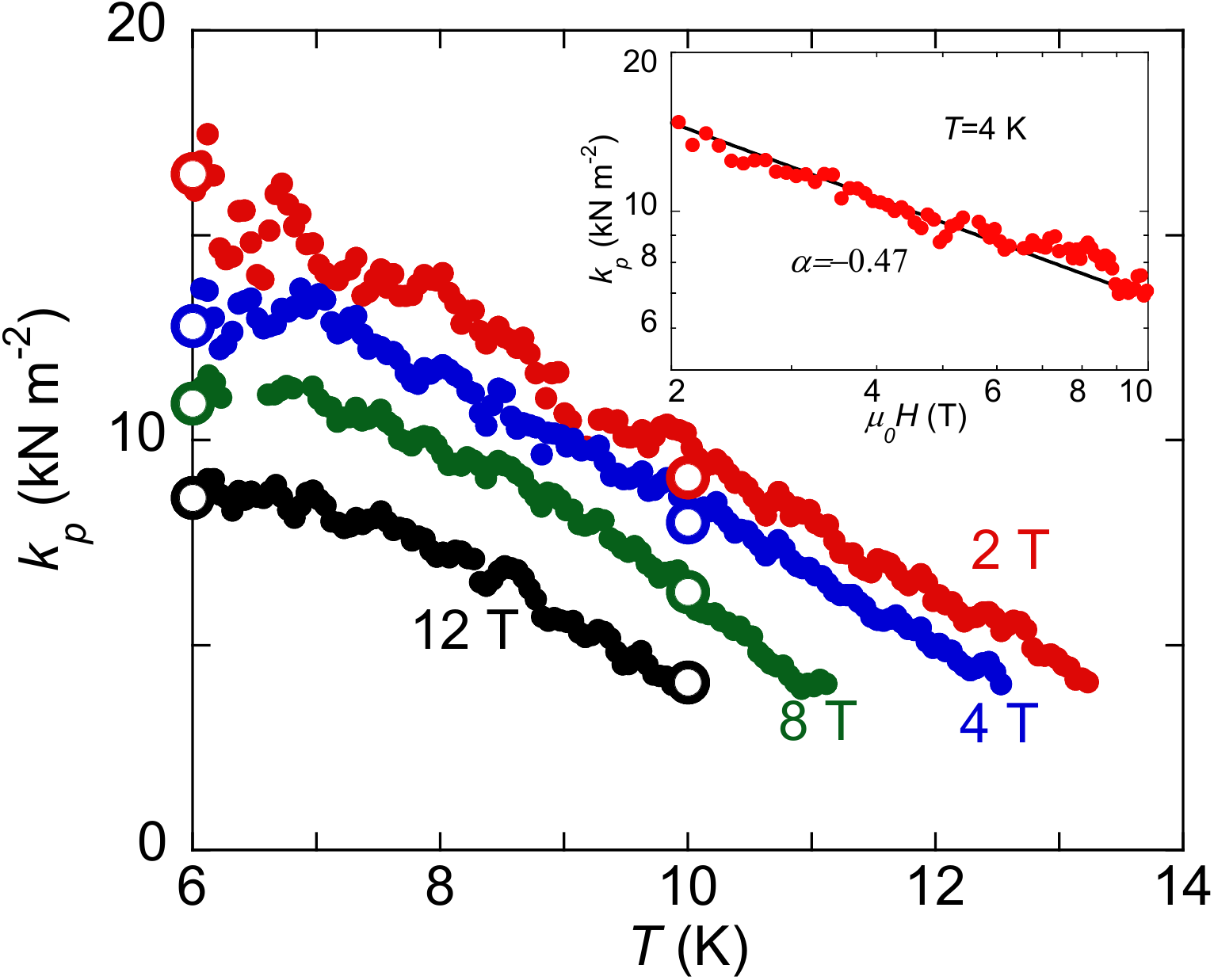}
\caption{Measured pinning constant $k_p(H,T)$ in FC condition at {2~T, 4~T, 8~T, 12~T}. The sparse empty cyrcles come from the ZFC measurements. In the inset $k_p(H)$ at 4~K is shown in a log-log plot to highlight the power dependence $k_p\propto H^{\alpha}$, with $\alpha=-0.47$, typical of the collective pinning regime.}
\label{fig:kp}
\end{figure}

The pinning constant, shown in \figurename~\ref{fig:kp}, is obtained by combining Eq.s~(\ref{eqn:rhovm1}) and (\ref{eqn:rhovm2}) and Eq.~(\ref{eqn:rhoGR}), $k_p=2\pi\Phi_0B\nu_p/\rho_{ff}$.
As presented in Sec.~\ref{sec:theory} this parameter in the limiting case of rigid vortices is a measure of the steepness of the pinning potential wells. 
The obtained $k_p$ correctly decreases when the temperature and the magnetic field are increased due to a reduction of the pinning efficiency. \figurename~\ref{fig:kp} shows that even at ${\mu_0H=8}$~T and ${T\sim8}$~K, ${k_p>10}$~kN/m\textsuperscript{2}. 
%
{This value indicates an enhanced pinning efficiency in \NbSn\ as compared to that of Nb films. In fact, in the latter, the $k_p$ literature value is assessed to be about an order of magnitude smaller than that of \NbSn\ at ${t=0.5}$ in a 40~nm thick film \cite{janjuvsevic2006microwave} and even smaller in a 30~nm thick film at ${t=0.86}$ \cite{Silva2011,Pompeo2012}.}
{A $k_p$ value similar similar to that of \NbSn\ was} also observed on pristine bulk MgB\textsubscript{2}, where at 1~T and  at 10~K $k_p\sim11$~kN/m\textsuperscript{2}\cite{alimenti2020microwave}. 
Higher $k_p$ are observed in cuprates, 
e.g. $k_p\simeq75$~kN/m\textsuperscript{2} at $t=0.5$ and $\mu_0H=0.5$~T in 100~nm \YBCO\ thin film added with BaZrO$_3$ inclusions  \cite{torokhtii2016microwave}, and $k_p$ up to 100~kN/m\textsuperscript{2} attained even at much higher $t\sim0.74$, at $\mu_0H=0.5$~T, in 200~nm \YBCO\ thin film added with Ba$_2$YNbO$_6$+Ba$_2$YTaO$_6$ inclusions \cite{Bartolome2020, Torokhtii2020}.

The temperature dependence $k_p(T)$ is shown in \figurename~\ref{fig:kp}: $k_p$ decreases steadily with the temperature indicating that no matching-field effects take place. The temperatures for which $k_p=0$, corresponding to the complete vanishing of the pinning effect, are obtained through a linear extrapolation of the high temperature region of the curves in \figurename~\ref{fig:kp}. The corresponding  points are reported on the phase diagram of \figurename~\ref{fig:Bc2MG}. These points mark the depinning line as obtained by the microwave technique. As it can be seen, the complete flattening of the pinning potential arises very close to $H_{c2}$.

The ZFC measured $k_p(H)$ at 4~K is shown in the inset of \figurename~\ref{fig:kp} to follow the power law dependence $k_p\propto H^{\alpha}$ with $\alpha=-0.47$. This behavior is expected in the collective pinning regime where for conventional superconductors one expects $\alpha=-0.5$ \cite{golosovsky1996high,campbell1971interaction,gittleman1966radio}. In fact, this field dependence indicates that, even at low temperature, vortices in \NbSn\ are not individually pinned but a bunch of vortices is bounded  around weak pins, thus the vortices concentration is higher than that of the pinning centres. We indicate this pinning regime as \textit{collective pinning} according to \cite{golosovsky1996high}. In this configuration the fluxons interact with each other, and thus the pinning properties are strongly dependent on the fluxons density and the pinning strength decreases with the field.  
In this regime $k_p$ is no more a direct measure of the single pinning centre strength but it is a statistical average of the contribution given by several pinning centres and vortices. This means that in principle there is still room of improvement for enhanced  $k_p$ values in \NbSn\ samples engineered for high field and high frequency applications (e.g. RF cavities for dark matter research \cite{di2019microwave}). 
{In fact an upper limit for $k_p$ can be estimated in the single-vortex pinning regime by assuming vortices individually pinned by cylindrical defects of diameter $2\xi$, being $\xi$ the coherence length, oriented parallel to the applied magnetic field.}
In this case the condensation energy (per unit length) in the vortex core $\frac{1}{8}\mu_0H_c^2\xi^2$, with  $H_c$ the thermodynamic critical field, is equal to the maximum pinning elastic energy (per unit length) $\frac{1}{2}k_p\xi^2$. Hence, the maximum $k_p^{max}\approx 0.25\mu_0H_c^2$ \cite{golosovsky1996high} can be assessed in this ideal core pinning configuration. Using the literature value $\mu_0H_c(0)\sim0.52$~T \cite{godeke2006nb3sn} for stoichiometric \NbSn, $k_p^{max}\sim50$~kN/m\textsuperscript{2} is obtained. This $k_p$ upper limit is near to that measured on 100~nm film pristine YBCO \cite{torokhtii2016microwave}.


\subsubsection{Depinning frequency.}\label{sec:nup}
\begin{figure}[htbp]
\centering
\begin{subfigure}[t]{1\linewidth}
\centering
\includegraphics[width=0.6\linewidth]{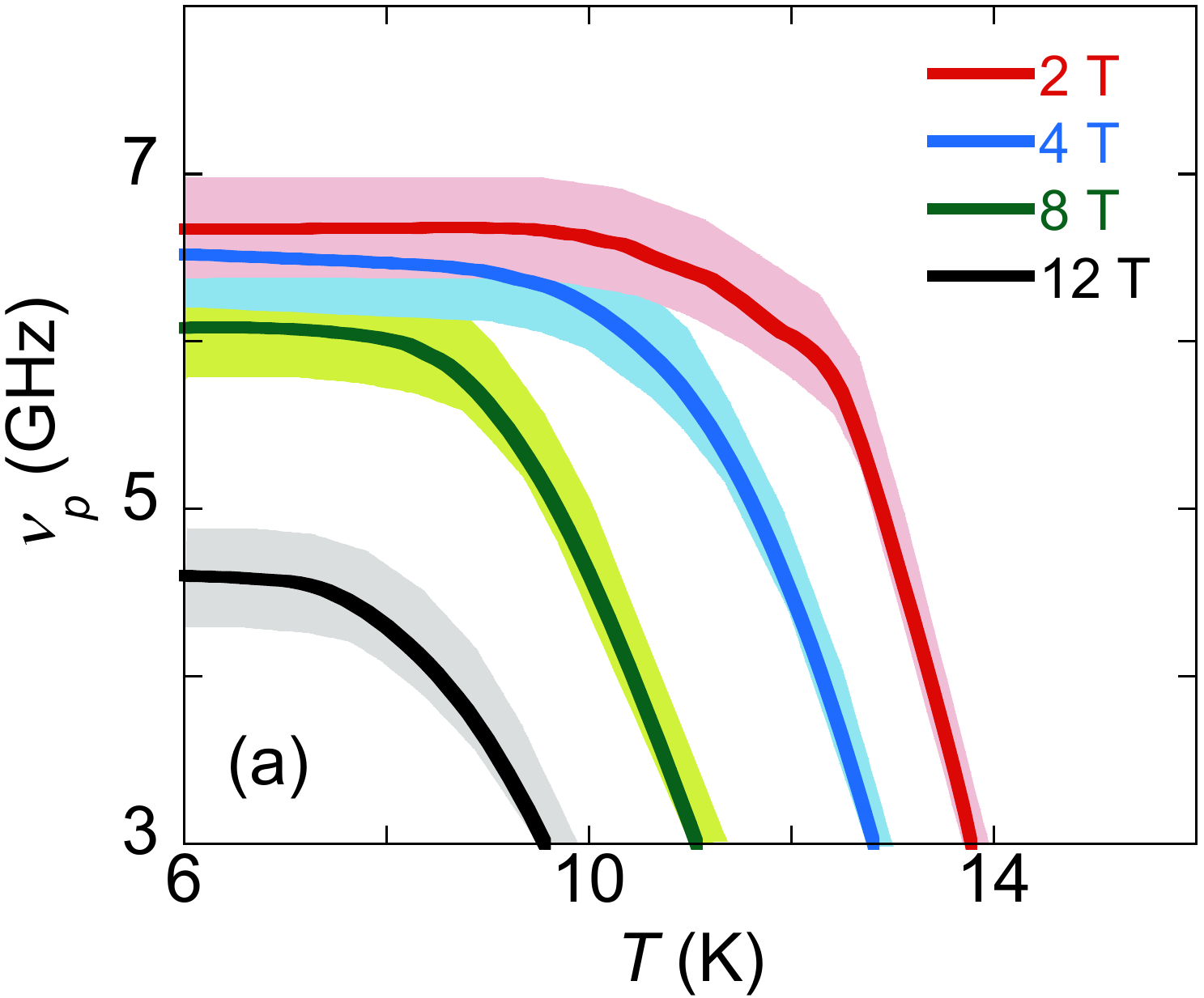}
\caption{}
\label{fig:nuP}
\end{subfigure}
\begin{subfigure}[t]{1\textwidth}
\centering
\includegraphics[width=0.6\linewidth]{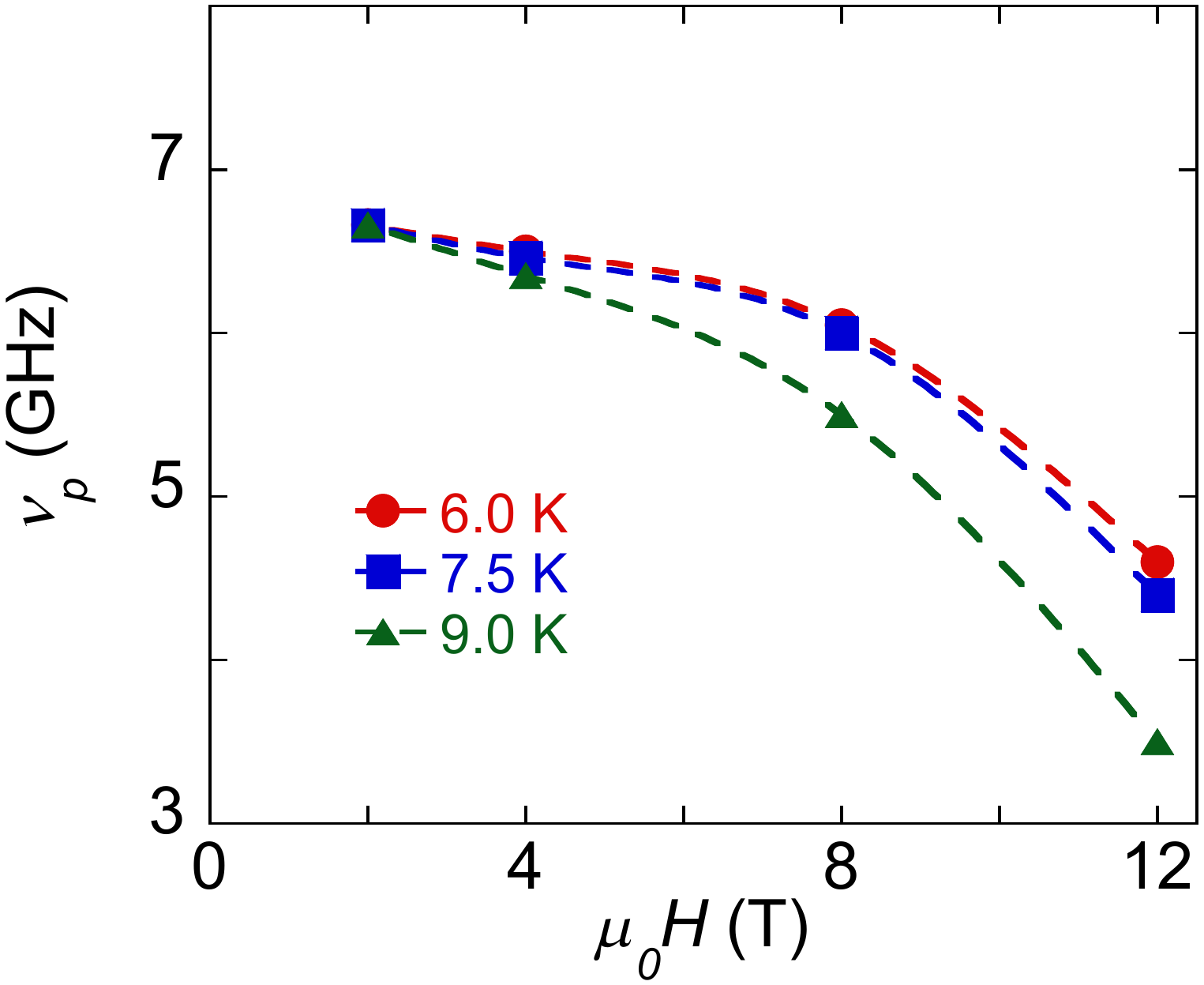}
\caption{}
\label{fig:nuP_B}
\end{subfigure}
\caption{(a) The depinning frequency $\nu_p$ measured in FC condition at {different fields (i.e. 2~T, 4~T, 8~T, 12~T)} obtained with the GR model. The shown data are smoothed and the standard deviation of the data scattering represented by the shadowed areas. The full dots are obtained from the ZFC measurements. (b) The depinning frequency $\nu_p$ dependence to the applied magnetic field $\mu_0H$ at {6.0~K, 7.5~K, 9.0~K}. The dashed line is a guide for the eye.}
\label{fig:nup_sideByside}
\end{figure}

Finally, the depinning frequency $\nu_p=k_p/(2\pi\eta)$ is discussed in this section. We show in \figurename~\ref{fig:nuP} the depinning frequency $\nu_p$ measured in FC condition at ${\mu_0H=\{2,4,8,12\}}$~T and obtained with Eq.~(\ref{eqn:rhoVM}). 
We can see that it is almost constant at low enough temperature (i.e. for ${T/T_{c2}<0.7}$) and it sharply decreases approaching the depinning line as described in Sec.~\ref{sec:pinnConst}. In \figurename~\ref{fig:nuP_B} the  $\nu_p(H)$ field dependence at $T=\{6,\,7.5,\,9\}$~K is shown. 
We note that at the lower $T$ $\nu_p(H)$ starts to decrease above $\mu_0 H=4$~T, while at lower fields it tends to saturate at $\nu_p\sim6.5$~GHz.
The measured values are quite large also at high fields, $\nu_p>4$~GHz at 12~T and low T which is larger than that exhibited by thin Nb films. 
It is known that $\nu_p$ is strongly dependent on the sample thickness in Nb films: 
$\nu_p\sim20$~GHz in 10~nm Nb film at 0.2~T and 5~K \cite{janjuvsevic2006microwave},  $\nu_p\sim5$~GHz in 60~nm Nb film at 0.6~T \cite{Silva2011}  and it falls to 1~GHz  for 160~nm films in 0.2~T and 5~K \cite{janjuvsevic2006microwave}. 
{In Nb the increase of $\nu_p$ with the lowering of the film thickness was attributed to the dominant effect of the surface pinning centres \cite{janjuvsevic2006microwave}. This effect is masked in thicker samples due to the increased volume interested by the weaker volume pinning in Nb \cite{janjuvsevic2006microwave}. Moreover, it is well known that the main contribution to pinning in \NbSn\  \cite{hanak1967flux,scanlan1975flux,shaw1976grain,zerweck1981pinning} as in other intermetallic compounds \cite{nembach1969electron,tanaka1976microstructures} and metals as Nb \cite{dasgupta1978flux,conrad1967microstructure,santhanam1976flux} is given by the grain boundaries and that the pinning efficiency is inversely proportional to the average grains size. Since in \NbSn\ the grain size can be reduced by lowering the sample thickness \cite{shaw1976grain}, it is reasonable to expect that for thin \NbSn\  samples $\nu_p$ could reach very high values.} This opens the possibilities to interesting RF applications of \NbSn\ films also in presence of high magnetic fields. Moreover, assuming a \NbSn\ sample engineered with a sufficiently high defects density to firmly remain in the single vortex pinning regime, from the previously calculated $k_p^{max}$, a theoretical upper limit $\nu_p^{max}\sim16$~GHz can be expected in bulk samples. 

Considering other superconducting materials, it comes out that the obtained values at 4~T are comparable with those measured in a MgB$_2$ thin film in the same H-T region \cite{Silva2016}).
On the other hand, it must be noticed that FeSe\textsubscript{0.5}Te\textsubscript{0.5} and \YBCO\ performances are still far, in fact at 12~K and 0.6~T ${\nu_p\sim22}$~GHz and $>40$~GHz, respectively in 300~nm and 240~nm thick FeSe\textsubscript{0.5}Te\textsubscript{0.5} films  \cite{pompeo2020microwave, pompeoFeSeTe2020} while ${\nu_p\sim50}$~GHz in 100~nm thick \YBCO\ films at 72~K \cite{torokhtii2016microwave}. For a more complete comparison, thin \NbSn\ films should be characterized in the same conditions to experimentally verify the increase of ${\nu_p}$ with the reduction of the sample thickness. Despite of this, from this study, it is shown that bulk \NbSn\ could remain a good choice for applications that work at not too high frequencies (e.g. radio frequency cavities for axions detection \cite{RevModPhys.75.777}) and for which the use of a metallic and wieldy material is an important requirement.

\subsubsection{Evaluation of the thermal creep contribution.}
We complete this work by providing an estimate of the thermal creep contribution to the evaluation of the vortex parameters through a statistical analysis according to \cite{pompeo2008reliable}. We derive the maximum creep factor $\epsilon_{max}$ and the lower limit for the activation energy $U_{0,min}$. We then derive a confidence interval for the characteristic frequency $\nu_c$ (we recall that when creep is taken into account, the characteristic frequency is no longer $\nu_p$, but $\nu_c$, see Eq.~(\ref{eqn:rhoVM}) and Eq.~(\ref{eqn:rhoGR})). 


The maximum creep factor ${\epsilon_{max}=1+2r^2-2r\sqrt{1+r^2}}$, with ${r=\rho_{vm}''/\rho_{vm}'}$, is obtained from analytical constraints \cite{pompeo2008reliable}. Then, the corresponding minimum activation energy is determined with the CC model, in the scenario of a periodic pinning potential, since $\epsilon=(I_0(U_0(T,B)/(2k_BT)))^{-2}$ \cite{coffey1991unified} with  $I_0$ the modified Bessel function of first kind.



\begin{figure}[htbp]
\centering
\includegraphics[width=0.6\textwidth]{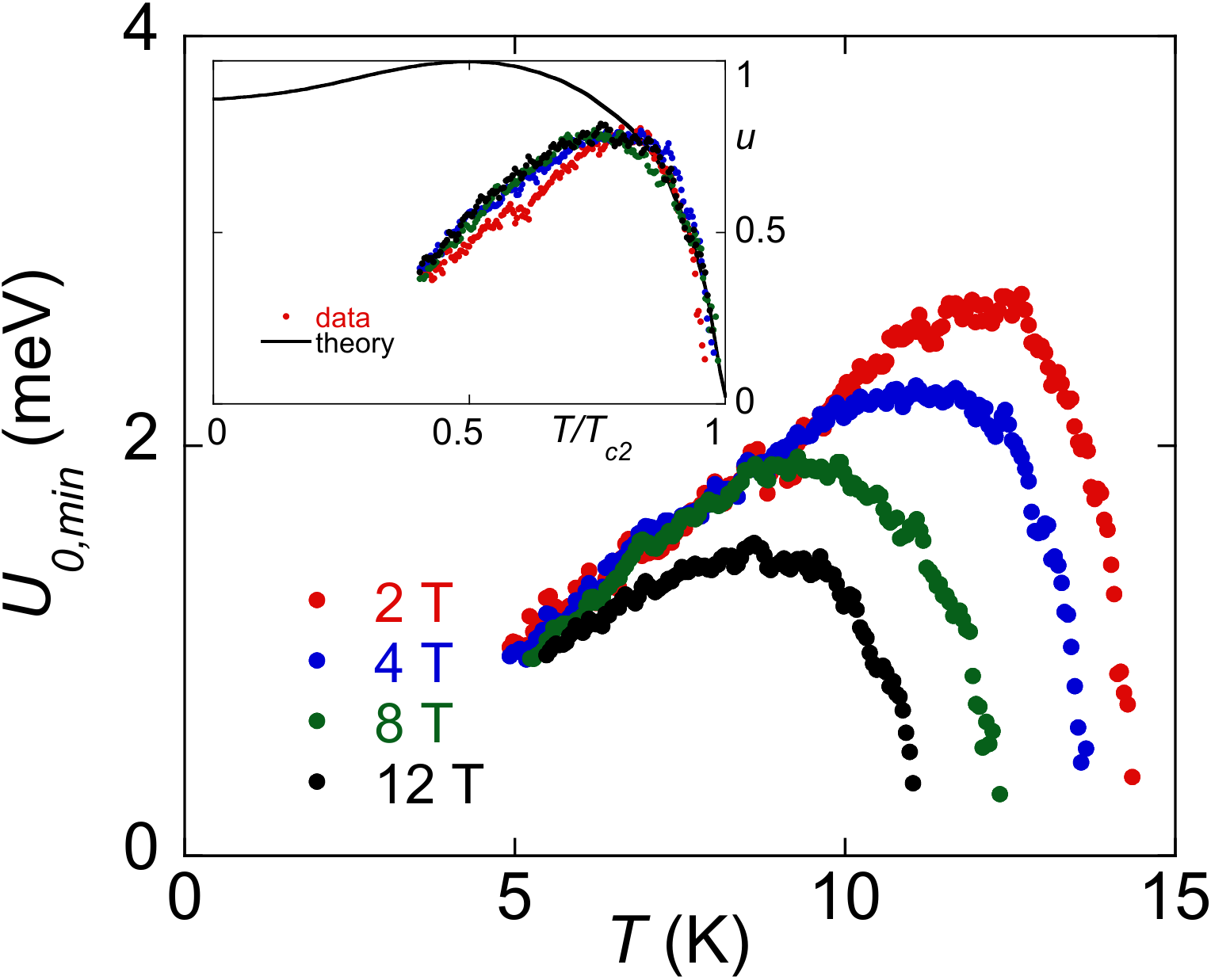}
\caption{Lower limit of the creep activation energy $U_{0,min}(T,H)$ in FC at ${\mu_0H=\{2,4,8,12\}}$~T. We show an almost perfect scaling of the $U_{0,min}(T,H)$ curves in the inset. The continuous line in the inset is the fit realized with the normalized $U_0\propto H_c^2\xi^n$. }
\label{fig:actEnergy}
\end{figure}

The lower limit for the activation energy $U_{0,min}(T,H)$ is shown in \figurename~\ref{fig:actEnergy}. 
The data show a non-monotonic temperature dependence with a peak height which becomes smaller, and moves at lower $T$, as the field is increased. This behavior is expected for $U_0$ since at higher $H$ values the pinning becomes less effective. Since $U_0\propto H_c^2\xi^n$, then the $U_0$  temperature dependence can be evaluated from ${H_c=\Phi_0/(\mu_0\sqrt{8}\pi\lambda\xi)}$ and using the BCS $\lambda$ previously used for the data analysis and $\xi=\sqrt{\Phi_0/(2\pi\mu_0H_{c2})}$ \cite{tinkham1996introduction} with the measured $H_{c2}$ from \figurename~\ref{fig:Bc2MG}. The $n=0,1,2$, or 3 parameter depends on the relevant length scale for the pinning energy: it indicates the dimensions of the correlated volume of the fluxons bunch that is thermally activated. From the theoretical behavior for $U_0$, the observed non-monotonic trend can be obtained only with $n=3$. Keeping this value for $n$, a tentative comparison between the theoretical curve and the experimental data, arbitrarily scaled with the constraint $U_{0min}<U_0$ for each $T$, is reported in the inset of \figurename~\ref{fig:actEnergy}. Assuming that the temperature behavior $U_{0,min}(T)$ reflects $U_0(T)$, $n=3$ is an indication that in this sample the vortices correlated volume has a length scale $\xi$ along the three spatial directions.
It must be noticed that the peak in $U_{0,min}$ is narrower than what expected from the theory and also that the lower temperature $U_{0,min}$ dependence does not saturate to a finite value but it is a linear function of $T$.
The narrow peak and the increase of $U_0$ with the temperature was already observed in other superconductors \cite{keller1990magnetic, miu2004nonmonotonic,leo2015vortex, pompeoFeSeTe2020}. 
This discrepancy with respect to the theory was justified introducing pinning models that included junctions and non-homogeneities \cite{griessen1990resistive,feigel1989theory,chikumoto1992flux,gurevich1991transient,keller1990magnetic}. 

To evaluate the impact of finite flux creep on the estimate of the vortex parameters, one should know the statistical distribution of the activation energies. Although we can set $U_{0,min}$ from the data, a full knowledge of the statistical distribution is not available. We then model the distribution of $U_{0}$ as a rectangular (uniform) distribution, and we seek for an estimate of the maximum (cutoff) $U_{0,max}$. 
The latter is determined consistently with the models used in this analysis. In particular, in the ideal case one can assume that the measured $k_p$ is not dependent to the fluxon displacement (i.e. that the pinning wells have perfectly parabolic profiles) and that the wells maximum width is $2\xi$. In this case the maximum elastic pinning energy is $U_{0,max}=\frac{1}{2}k_p\xi^2l$ where $l$ is the length of the effective pinning along the direction of the applied magnetic field \cite{golosovsky1996high}. 
According to the indication $n=3$ (\figurename~\ref{fig:actEnergy}), $l\sim\xi$, then we set $l\simeq\xi$. The coherence length is obtained from the previously determined $H_{c2}$ (\figurename~\ref{fig:Bc2MG}), $\xi=\sqrt{\Phi_0/(2\pi\mu_0H_{c2})}$ \cite{tinkham1996introduction} and from the GR model $k_p$ a first estimation of $U_{0,max}$ is obtained. Actually, since the creep is now taken into account, the $U_{0,max}$  estimation can be enhanced with a recursive approach: once $U_{0,max}$ is obtained from the GR $k_p$, it can be used to calculate the creep factor $\epsilon$ to be used in Eq.~(\ref{eqn:rhoVM}), thus a new $k_p$ can be obtained with the CC model. This in turns fixes $\epsilon$, and a refined value for $k_p$ can be evaluated from the measured data and the fixed $\epsilon$ with the CC model. In this way $U_{0,max}$ is evaluated several times until $(k_p^{i+1}-k_p^{i})/k_p^i<0.01$, with $i$ the iteration number (e.g. at $T=9$~K and $\mu_0H=2$~T the problem converges in 5 steps). The $U_{0,max}$ obtained from the last iteration is used for the statistical analysis now presented.


\begin{figure}
\centering
\includegraphics[width=0.6\textwidth]{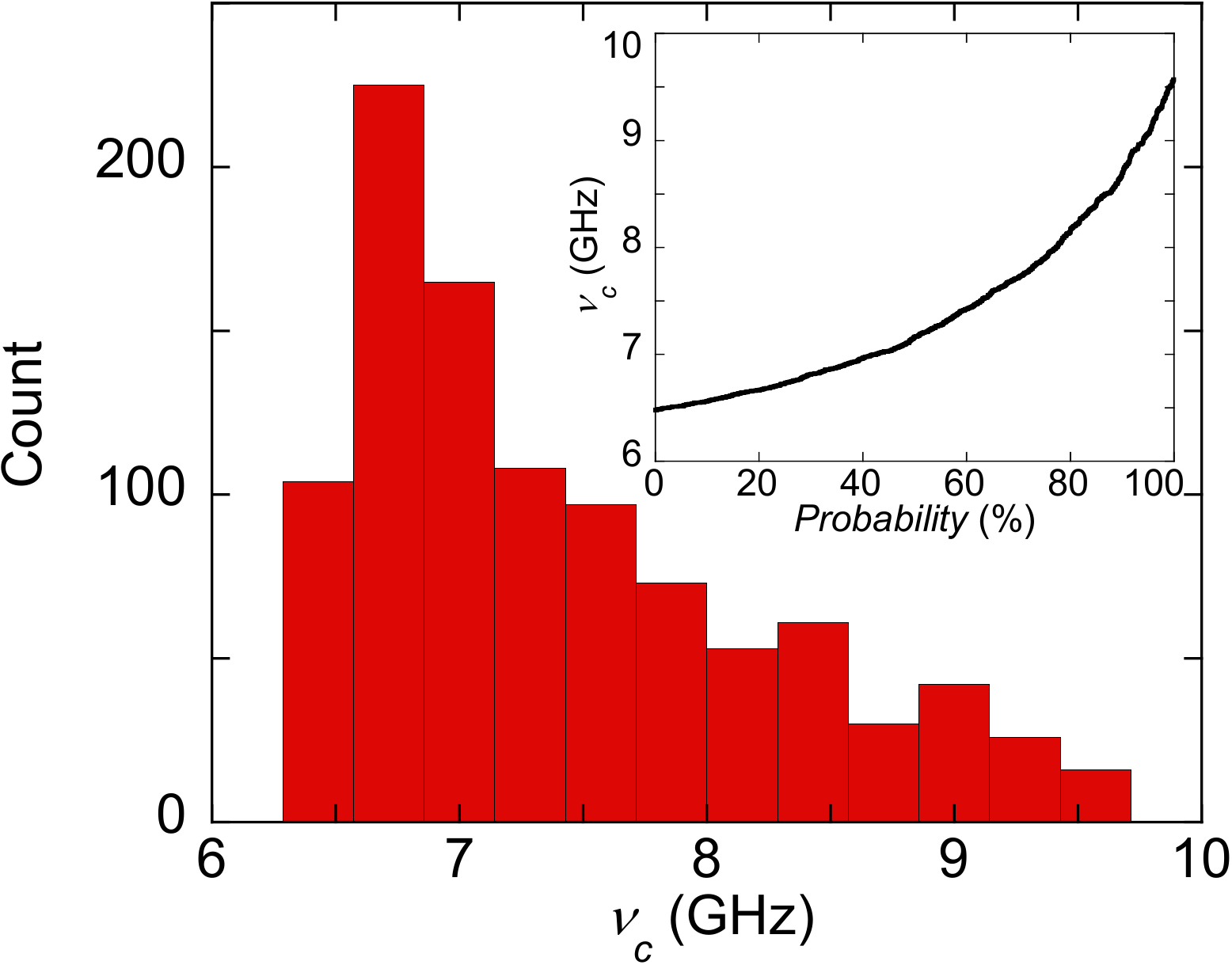}
\caption{Characteristic frequency $\nu_c$ distribution obtained by evaluating Eq.~(\ref{eqn:rhoVM}) with 1000 randomly chosen $\epsilon$ values with the statistical distribution described in the text. In the inset the percentage cumulative probability distribution. It shows that with a probability level of $90\,\%$ ${6.5<\nu_c/\mbox{GHz}<8.7}$ while ${6.5<\nu_c/\mbox{GHz}<7.7}$ at $68\,\%$.  }
\label{fig:nuP_distribution}
\end{figure}

With $U_0$ taken to follow a rectangular distribution between $U_{0,min}$ and $U_{0,max}$, we recalculate the pinning parameters using Eq.~(\ref{eqn:rhoVM}) with 1000 $\epsilon$ values randomly extracted from the $U_0$ distribution previously built. We show as an exemplary case the creep contribution evaluated on $\nu_c(t=T/T_c=0.5,\mu_0H=2\,\mbox{T})$. We focus on $\nu_c$ since it is the cross-over frequency between the low frequency vortex elastic motion and the high frequency dissipative region. Thus in case of creep $\nu_c$ is the parameter of interest for technological applications. The characteristic frequency $\nu_c$ distribution, with its  cumulative probability distribution, is shown in \figurename~\ref{fig:nuP_distribution}.  The expected value $E[\nu_c]_{CC}=7.4$~GHz is about $20\,\%$ larger than the numerical value for $\nu_p$ within the GR model in the same condition. At $t=0.5$ and $\mu_0H=2\,\mbox{T}$, we evaluated with the CC model also the $\rho_{ff}$ distribution obtaining ${(E[\rho_{ff}]_{CC}-\rho_{ff,GR})/\rho_{ff,GR}\sim3\,\%}$. Whereas, at $t=0.5$ and $\mu_0H=12$~T, where the creep phenomenon is more intense, the discrepancy between the CC and GR valued parameters is about  $35\,\%$ on $\nu_c$ and $6\,\%$ on $\rho_{ff}$.

As expected from \cite{pompeo2008reliable}, one can notice that the GR evaluation gives a lower boundary for $\nu_p$, $\rho_{ff}$ and $k_p$. In particular, even if neglecting the thermal creep brings to an underestimation of the material characteristic frequency $\nu_c$, a drastic difference is not expected from that obtained with the more complete CC model. Thus, it means that for the design of perspective RF/high field applications of \NbSn\ the $\nu_p$ shown in Sec.~\ref{sec:nup} can be treated as the worst RF performance of \NbSn\ but as an indication of a more realistic value of $\nu_c$, the expected values shown in this section can be used. {Whereas, regarding $\rho_{ff}$} it can be assessed that the GR determination can be considered reliable since even at the highest creep rate its shift is modest (i.e. $\leq6\,\%$).

{\section{Vortex parameters comparison with other SCs}\label{sec:Comparison}
In this section a brief comparison between the mixed-state microwave properties of \NbSn\ and those of other technologically interesting superconductors (i.e. \MgB\ and \YBCO\ ) is provided.}

\begin{figure}
\centering
\includegraphics[width=0.6\textwidth]{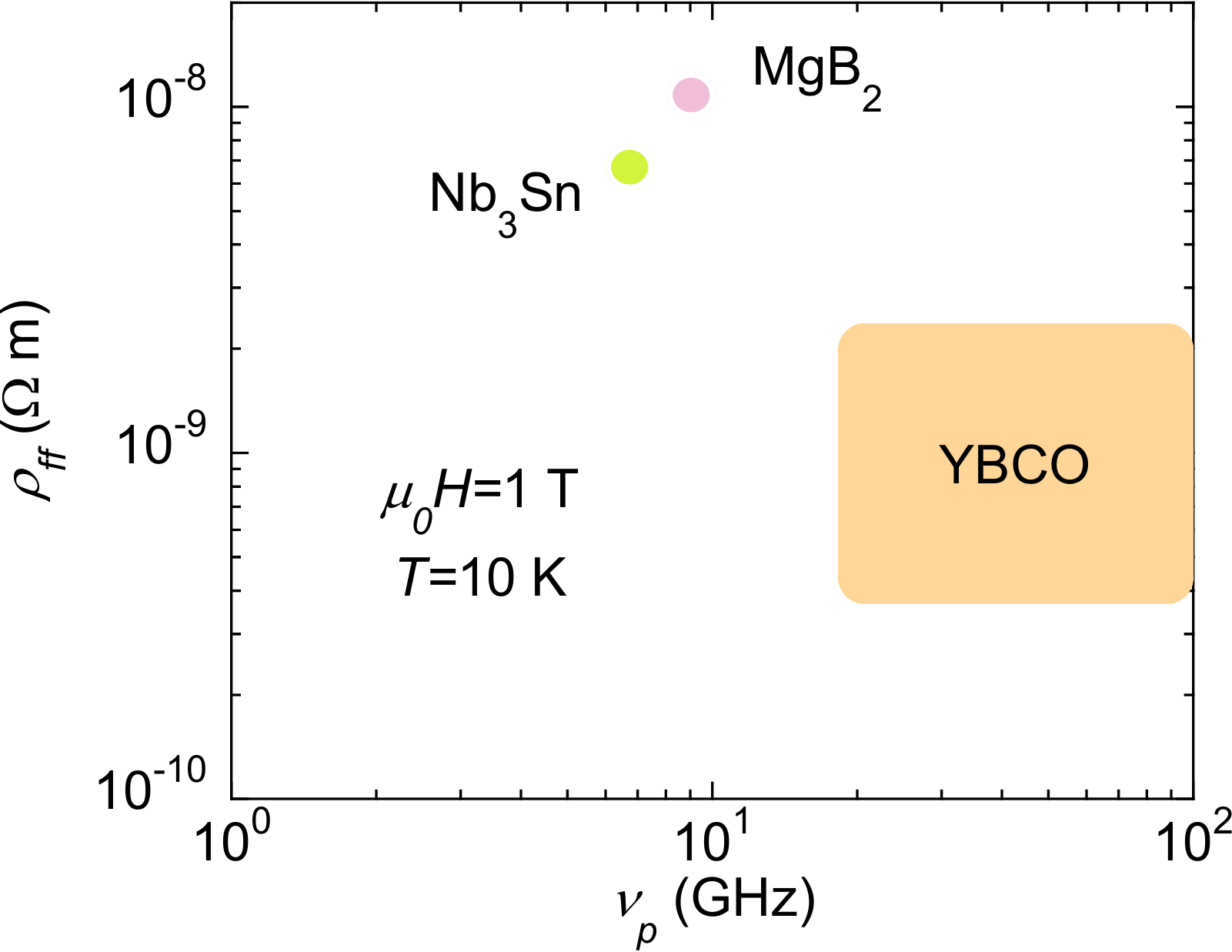}
\caption{{Comparison between the \mw\ vortex parameters ($\rho_{ff}$ and $\nu_p$) of \NbSn\, \MgB\ and \YBCO\ at $\mu_0H=1$~T and $T=10$~K.}}
\label{fig:ConfrontoNbSn_MgB_YBCO_APPL}
\end{figure}

\begin{figure}
\centering
\includegraphics[width=0.6\textwidth]{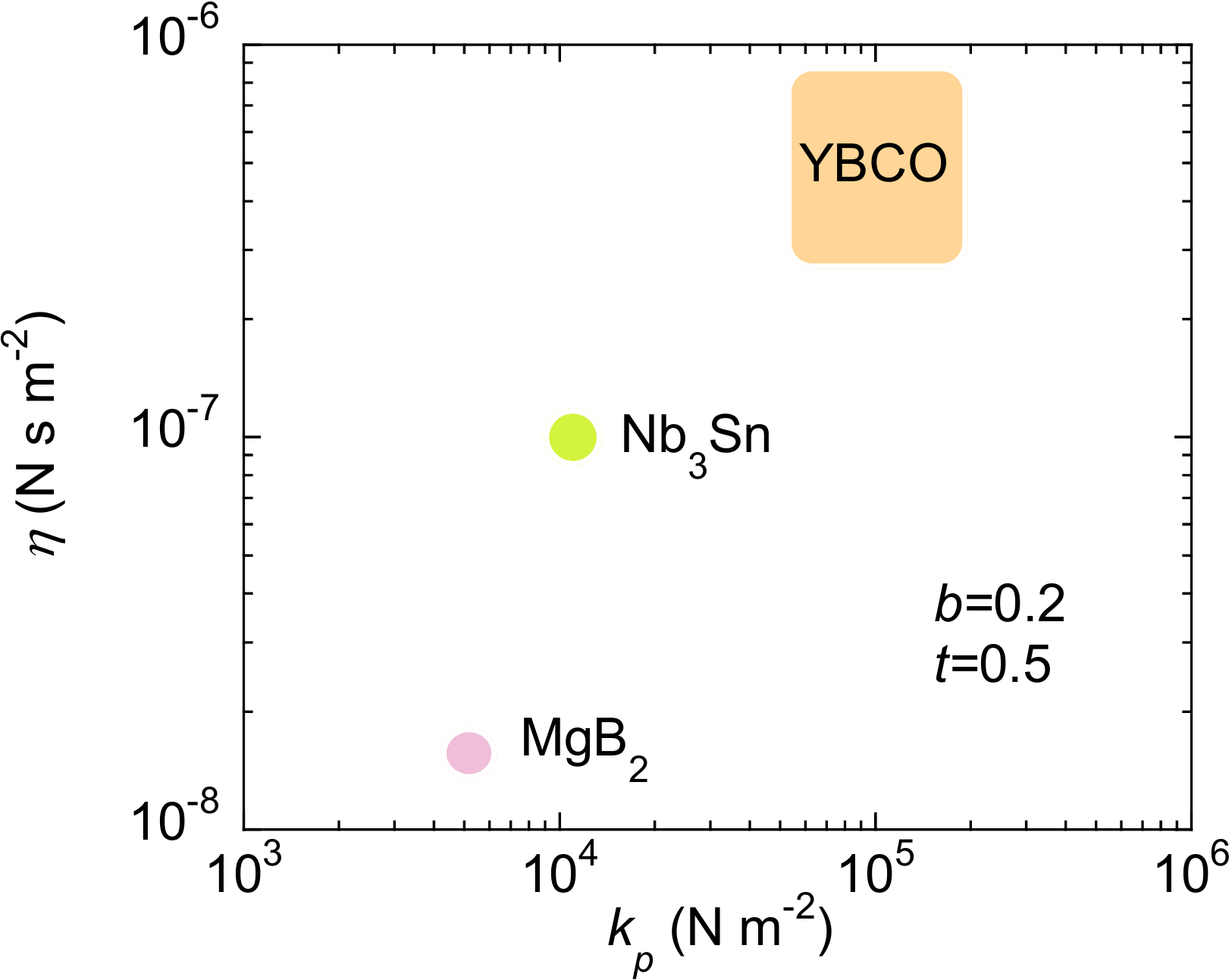}
\caption{{Comparison between the pinning constant $k_p$ and the viscous drag coefficient $\eta$ of \NbSn\, \MgB\ and \YBCO\ at $t=0.5$ and $b=0.2$.}}
\label{fig:ConfrontoNbSn_MgB_YBCO_PHYS}
\end{figure}

{In order to keep the comparison as meaningful as possible, we compare data obtained mainly by our group on \MgB\ and YBCO with the same technique and with the use of the same physical model (e.g. the GR model). The \MgB\ data were obtained   on  bulk  pristine and doped samples at $16.5$~GHz and $26.7$~GHz up to $1.2$~T. Further details on the \MgB\ characterization are shown in \cite{alimenti2020microwave}. For what concern the YBCO, the parameters used for the comparison come from  several literature results on thin films \cite{pompeo2013anisotropy,Torokhtii2020,Bartolome2020,7433952,torokhtii2016microwave}, commercial coated conductors \cite{torokhtii2016measurement,romanov2020high} and single crystals \cite{tsuchiya2001electronic}. The vortex parameters are linked to each other by $\nu_p=k_p/(2\pi\eta)$ and $\eta\propto\rho_{ff}^{-1}$ Eq.~(\ref{eqn:rhof_rhon}). It is then useful to investigate the parametric plots as reported in Figures~\ref{fig:ConfrontoNbSn_MgB_YBCO_APPL} and \ref{fig:ConfrontoNbSn_MgB_YBCO_PHYS}.  In \figurename~\ref{fig:ConfrontoNbSn_MgB_YBCO_APPL} the comparison at $\mu_0H=1$~T and $T=10$~K is shown on the plane $\rho_{ff}-\nu_p$. From this, one can notice that \NbSn\ shows the lowest $\nu_p$ of the three SC materials. Despite of this, it must be noticed that the large \MgB\ $\nu_p$ comes from its particularly large $\rho_{ff}$ as shown in \cite{alimenti2020microwave}. Thus, despite the larger $\nu_p$, the \mw\ losses are smaller in \NbSn\, with respect to those in \MgB, even above  $\nu_p$. In particular, in \MgB\ $\rho_{ff}$ exhibits a non conventional Bardeen-Stephen behavior \cite{shibata2003anomalous,alimenti2020microwave} due to the presence of the weak superconductive $\pi-$band. This makes \MgB\ advantageous for \mw\ applications in the mixed state only in particular conditions, e.g. below the field values for which the smaller gap is suppressed. Finally, as shown in the previous sections, YBCO performances are still better than those of metallic SCs. On the other hand, the practical use of YBCO in large-scale \mw\ applications like cavities is hindered by the difficulties in the deposition on continuous, and possibly non-planar, surfaces. It must be noticed that in \NbSn\ $\nu_p$ is high enough for applications like the dark matter cavity detectors \cite{RevModPhys.75.777}, thus despite its lower $\nu_p$ it retains its importance for \mw\ applications.}

{From the technological point of view \figurename~\ref{fig:ConfrontoNbSn_MgB_YBCO_APPL} shows a useful comparison of the \mw\ most interesting parameters of these SCs at fixed working conditions (i.e. at $\mu_0H=1$~T and $T=10$~K). However, it can be useful to compare the different SCs also with a more physical approach, evaluating $k_p$ and $\eta$ at the same reduced temperature $t=T/T_c$ and field $b=B/B_{c2}$. This comparison is shown in \figurename~\ref{fig:ConfrontoNbSn_MgB_YBCO_PHYS}. The higher $k_p$ in YBCO is caused by the single fluxon pinning in this kind of SC \cite{golosovsky1996high}, while in both \NbSn\ and \MgB\ it was shown that the vortex system is in the collective pinning regime \cite{alimenti2020microwave}. However, it must be noticed that the $k_{p,max}$ here theorized for \NbSn\ corresponds to the lower limit for $k_p$ in YBCO. Thus, in theory if it would be possible to optimize \NbSn\ with artificial pinning centres effective at microwaves, the high frequency performances of \NbSn\ could be expected to be near that of YBCO. Finally, the viscous drag coefficient $\eta=\Phi_0B/\rho_{ff}\propto<\tau>$ shows that the quasi-particles scattering time $\tau$ in the fluxons cores is particularly reduced in \MgB\ because of the high normal carriers density coming from the suppressed $\pi$-band \cite{sarti2005dynamic,alimenti2020microwave}. In \NbSn\ $\eta$ is about 3 times smaller than the lower $\eta$ value in YBCO, thus even if for $k_p$ and $\nu_p$ there is still room for improvement, the \NbSn\ microscopic properties would still limit the high frequency dissipation in this material.}

{The comparison between the \NbSn, \MgB\ and \YBCO\ \mw\ vortex parameters shows that \NbSn\ exhibits intermediate performances between those of the other two SCs. However, the possibility of increasing $k_p$, with artificial pinning centres optimized to work at \mw\ in order to extend the single pinning regime in \NbSn, can be particularly interesting to obtain a metallic superconductor with improved \mw\ pinning characteristics, close to that of YBCO.}

\section{Summary}
A polycrystalline bulk \NbSn\ sample was characterized at 15~GHz in order to study the high-frequency vortex motion in high magnetic fields up to 12~T. The measurements were performed with a dielectric loaded resonator both in field cooling  and zero field cooling  conditions. 
The obtained normal state material parameters matched with the literature values.
The upper critical field $H_{c2}(T)$ was evaluated at the higher temperatures directly up to 12~T while at the lower temperatures $H_{c2}(T)$ was obtained by the scaling of the flux flow resistivity $\rho_{ff}$ in a self-consistent way: the scaling procedure based on the Bardeen-Stephen (BS) model \cite{PhysRev.140.A1197} $\rho_{ff}/\rho_n\sim H/H_{c2}$ gave a $H_{c2}(T)$ in agreement with the expected, from literature \cite{godeke2005upper}, Maki-de~Gennes (MG) behavior \cite{maki1964magnetic,de1964behavior} confirming both the validity of the scaling itself and the conventional BS behavior of bulk\NbSn.
The \NbSn\ depinning frequency $\nu_p$ reached rather high values, above 4~GHz even at 12~T and low $T$, indicating that \NbSn\ is suitable for radio frequency low loss applications  up to few GHz in bulk form.
Since $\nu_p$ strongly decreases with the film thickness, as shown in Nb, higher $\nu_p$ values can be expected in \NbSn\ thin films. 
The pinning constant $k_p(T,H)$ was found to decrease when the temperature and the field are increased due to the reduction of the pinning efficiency. In particular, a field dependence typical of the collective pinning scenario (i.e. $k_p\propto H^{-0.5}$) was shown. Despite the collective pinning, $k_p>10$~kN~m\textsuperscript{-2} for $H\leq8$~T and $T\leq8$~K which is about 10 times greater the values found in thin Nb films \cite{janjuvsevic2006microwave}. An estimation of the maximum $k_{p,max}\sim50$~kN/m\textsuperscript{2}, corresponding to single-vortex core pinning, shows that an alternative path to higher $\nu_p$ (from higher $k_p$) might arise from appropriate defect engineering.
In fact, only in the last years the necessity to operate at high frequencies and high magnetic fields emerged, while no particular material studies were undertaken for optimizing the superconductive properties in these harsh working conditions. 

Finally, we provided an analysis of the impact of the thermal activation on the vortex motion parameters through a statistical method and using the Coffey-Clem (CC) model \cite{coffey1991unified}. The creep activation energy $U_0$ distribution was modeled with a uniform distribution with upper and lower bound estimated consistently with the models and results already obtained.
Although the results obtained with the Gittleman-Rosenblum (GR) model \cite{gittleman1966radio} (assuming negligible creep) represent a lower limit for both $\rho_{ff}$ and $\nu_c$, we obtained that $\rho_ {ff}$, evaluated with the GR model, can be considered a reliable determination while for $t=0.5$ the expected value $E[\nu_c]_{CC}\sim1.2\nu_{p,GR}$ at 2~T and $E[\nu_c]_{CC}\sim1.4\nu_{p,GR}$ at 12~T.

This work represents, to our knowledge, the first report of the microwave response in \NbSn\ at high fields. The results here obtained are encouraging for the use of \NbSn\ in RF in high fields, although further optimization of the pinning can be needed for the specific requirements  of high-frequency applications.


\section*{Acknowledgements}

The authors warmly thank Carmine Senatore for useful discussions. 

\section*{References}

\bibliographystyle{iopart-num}
\providecommand{\newblock}{}

\end{document}